\documentclass[11pt,a4paper]{article}
\usepackage{jheppub}
\usepackage{epsfig}
\usepackage{latexsym}
\usepackage{amsfonts}
\usepackage{amsmath}
\usepackage{amsthm}
\usepackage{amssymb}
\usepackage{amsbsy}
\usepackage{multirow}

\newcommand{\CC}{\mathbb{C}} 
\newcommand{\RR}{\mathbb{R}} 
\newcommand{\ZZ}{\mathbb{Z}} 

\newcommand{\G}{\mathcal{G}}

\newcommand{\h}{\mathfrak{h}}

\def\calc         {{\cal C}}
\def\cald         {{\cal D}}

\def\calk         {{\cal K}}
\def\call         {{\cal L}}

\def\calq         {{\cal Q}}

\def\cals         {{\cal S}}

\def\calv         {{\cal V}}

\newsavebox{\uuunit}
\sbox{\uuunit}
    {\setlength{\unitlength}{0.825em}
     \begin{picture}(0.6,0.7)
        \thinlines
        \put(0,0){\line(1,0){0.5}}
        \put(0.15,0){\line(0,1){0.7}}
        \put(0.35,0){\line(0,1){0.8}}
       \multiput(0.3,0.8)(-0.04,-0.02){12}{\rule{0.5pt}{0.5pt}}
     \end {picture}}

\def\be{\begin{equation}}
\def\ee{\end{equation}}
\def\bea{\begin{eqnarray}}
\def\eea{\end{eqnarray}}
\def\bc{\begin{center}}
\def\ec{\end{center}}

\def\a{\alpha}
\def\b{\beta}
\def\h{\eta}

\def\G{\Gamma}
\def\d{\delta}
\def\e{\epsilon}
\def\D{\Delta}
\def\l{\lambda}

\def\k{\kappa}
\def\f{\phi}
\def\vf{\varphi}
\def\m{\mu}
\def\n{\nu}
\def\o{\omega}

\def\p{\pi}
\def\r{\rho}

\def\x{\xi}
\def\s{\sigma}

\def\t{\tau}

\def\bc{\bar{c}}

\def\sF{{{ F}\!\!\!\!\hskip.8pt\hbox{\raise1pt\hbox{/}}\,}}
\def\som{{{ \omega}\!\!\!\!\hskip.8pt\hbox{\raise1pt\hbox{/}}\,}}
\def\sJ{{{\rm J}\!\!\!\!\hskip.8pt\hbox{\raise1pt\hbox{/}}\,}}

\def\Im{\mathrm{Im}}
\def\Re{\mathrm{Re}}

\def\F{\Phi}
\def\pa{\partial}

\def\to{\rightarrow}
\def\nonu{\nonumber \\{}}
\def\half{{1 \over 2}}

\title{Unlocking the Axion-Dilaton in 5D Supergravity}
\author[a]{Joris Raeymaekers,}
\author[b]{Dieter Van den Bleeken}

\affiliation[a]{Institute of Physics of the ASCR, \\
Na Slovance 2, 182 21 Prague 8, Czech Republic.}
\affiliation[b]{Physics Department, Bo\u{g}azi\c{c}i University\\
 34342 Bebek / Istanbul, TURKEY}

\emailAdd{joris@fzu.cz}
\emailAdd{dieter.van@boun.edu.tr}

\abstract{We revisit supersymmetric solutions to five dimensional ungauged N=1 supergravity with dynamic hypermultiplets. In particular we focus on a truncation to the axion-dilaton contained in the universal hypermultiplet. The relevant solutions are fibrations over a four-dimensional K\"{a}hler base with a holomorphic axion-dilaton. We focus on solutions with additional symmetries
and classify Killing vectors which preserve the additional structure imposed by supersymmetry; in particular we extend the existing classification of solutions with a space-like U(1) isometry to the case where the  Killing vector is rotational.
We elaborate on general geometrical aspects  which we illustrate in some simple examples. We especially  discuss solutions describing the backreaction of M2-branes, which for example play a role in the   black hole deconstruction proposal for  microstate geometries. } 

\begin{document}
\maketitle

\section{Introduction}
Supergravity in 5 dimensions with 8 real supercharges plays an important role in our explorations of (quantum) gravity. Through type IIA/M-theory duality compactified on a Calabi-Yau threefold we can think of it as describing a strong coupling regime of 4 dimensional (super)gravity. One particular example where this is of use is in studying 4 dimensional extremal black holes, which were given a first quantum mechanical interpretation by Maldacena, Strominger and Witten \cite{Maldacena:1997de} using this connection between 4d/5d supergravity. More generally this is but one aspect of a particular AdS$_3$/CFT$_2$ correspondence, based on the maximally supersymmetric AdS$_3\times$S$^2$ vacuum solution to $N=1$ 5D supergravity.

Supersymmetric solutions to the theory have been much studied, and actually formally classified. The first step was made in the classic work \cite{Gauntlett:2002nw}, where the case of `pure' supergravity, i.e. the theory containing only the gravity multiplet, was analyzed. More recently this was extended to theories containing an arbitrary number of vector- and hypermultiplets in \cite{Bellorin:2006yr}. These general analyses are quite powerful and elegant, as they manage to simplify the Killing spinor equations into a small number of essentially geometric conditions. However in the presence of hypermultiplets one of these conditions,  the requirement that the hyperscalars  form a `quaternionic map' \cite{Bellorin:2006yr} is still quite hard to explicitly address, and leads to non-trivial and complicated consequences for the underlying geometry. These have only been worked out in a small number of explicit examples,  including \cite{Gutperle:2000sb},\cite{Gutperle:2000ve},\cite{Bellorin:2006yr},\cite{Emam:2006sr},\cite{Emam:2007qa}.

In this work we progress towards a more explicit understanding and formulation of such solutions with dynamical (i.e. non-constant) hypermultiplets. The main simplification that allows us to move forward is to make a consistent truncation in the hypermultiplet sector to the axion-dilaton part of the universal hypermultiplet. In doing so we will review parts of the general story, pointing out a few observations that went unnoticed in the literature.

Before we give a summary and overview of the paper, let us shortly mention the particular puzzle that motivated  this study. In \cite{Denef:2007yt}, a proposal was made for the  brane configuration representing a typical microstate geometry of the four dimensional D4/D0 black hole in type IIA compactified on a Calabi-Yau threefold. The configuration consists of a wrapped D6-brane and anti-D6 brane with worldvolume fluxes, surrounded by an ellipsoidal D2 brane. When lifted to five dimensional supergravity describing  M theory on a Calabi-Yau, this configuration becomes  a two-centered Taub-NUT system with an M2-brane which sources the axion-dilaton. In \cite{Levi:2009az} this configuration was argued to fit within a certain  ansatz for the metric and other fields.
However one can verify that this ansatz is incompatible with the classification of solutions with a spacelike  isometry in \cite{Bellorin:2006yr}.  The resolution to  this puzzle is, as we will see, is that in \cite{Bellorin:2006yr} the solution was assumed to be invariant under an  isometry of the   translational type, while the brane configuration of interest is instead  invariant under an  isometry of the   rotational type. We therefore need  to generalize the analysis of \cite{Bellorin:2006yr} to the case of a rotational
Killing vector, and we will indeed find that the results are compatible with the  ansatz of \cite{Levi:2009az}.
\subsection{Summary and overview}
In section \ref{charsec} we first review the structure of supersymmetric solutions to ungauged 5D N=1 supergravity following \cite{Bellorin:2006yr}. Such solutions always have at least one Killing vector and we will focus on the case where this  Killing vector is time-like. The metric can then be written as a time-like fibration over a 4D Euclidean manifold that is referred to as the {\it base}. When the hypermultiplets are constant this base is hyperk\"ahler, and when they are dynamical the base is more exotic, essentially almost quaternionic where the quaternionic structure is covariantly constant with respect to a scalar dependent connection. The hypermultiplets themselves are restricted to form a `quaternionic map'.

Inside the universal hypermultiplet, whose target space is ${SU(1,2) \over U(2)}$, there sits a complex scalar $\tau$ that parameterizes the subspace  ${SU(1,1) \over U(1)}$. We will refer to $\tau$ as the {\em axidilaton}, and its real and imaginary part can be identified with the Hodge dual of the totally external part of the M-theory 4-form and  the volume modulus of the Calabi-Yau respectively. The theory can be consistently truncated keeping $\tau$ as the only dynamical scalar in the hypermultiplet sector. Our main result of section \ref{charsec} is the analysis of the conditions for supersymmetry in this  truncated theory. We show that the base is a K\"ahler manifold and that the `quaternionic map' condition simply becomes the requirement that $\tau$ be holomorphic.
The backreaction of the axidilaton on the geometry is encoded in a differential equation for the K\"ahler potential, where the standard
Monge-Amp\`ere equation describing hyperk\"ahler geometry is now deformed by a source term depending on the the axidilaton, see (\ref{defMA}) below.

Even when all hypermultiplets are trivial and the base is hyperk\"ahler, not much is known about that base in the general case without isometries.
So to make progress, also when hypermultiplets are present, it will be useful to study the case where there is an additional symmetry on the base.
 Most solutions relevant for applications possess extra symmetries, and especially if one wants to make contact with 4D supergravity demanding a space-like U(1) isometry is natural.

 In section \ref{kilvsec} we work out how the presence of one or two Killing vectors further simplifies the solutions.
 We begin by classifying the isometries which preserve the additional structures imposed by supersymmetry.
As in the hyperk\"ahler case \cite{Boyer:1982mm,Gegenberg} one can distinguish between  holomorphic isometries of the `translational' and `rotational' type depending on whether or not they preserve the almost quaternionic structure. When the
axidilaton is turned on, there is a further distinction which  arises from  the SU(1,1) U-duality of the theory. Indeed, in order to have a symmetry  it is
sufficient for the axidilaton to be left invariant modulo a U-duality transformation when transported along the Killing vector.
 There are essentially four cases -- either $\tau$ is invariant ($I$) under such transport, or it transforms with a parabolic ($P$), elliptic ($E$) or hyperbolic ($H$) U-duality -- and we summarized how they constrain the functional form of $\tau$ in (\ref{taugensol}).

The simplest case ($I$), when $\tau$ is left invariant by the Killing vector, is the most straightforward and we discuss it in depth in section \ref{invKV}. We show how the sourced Monge-Amp\`ere equation reduces to a sourced SU($\infty$) Toda equation in the rotational case and to a sourced flat Laplace equation in the translational case. We also point out how one can simplify the remaining equations for the vector multiplets and highlight the rotational case, extending the analysis of \cite{Bena:2007ju} in the absence of hypermultiplets. To our surprise certain features, like the presence of a simple algebraic stability bound on the location of charge centers/branes, remain intact in the more involved solutions with a rotational Killing vector. In the last part of section \ref{invKV} we use an ansatz based on separation of variables that reduces the Toda equation underlying the geometry to a simpler sourced Liouville equation. This ansatz contains  solutions with a larger, non-abelian group of symmetries for the supergravity background, and we  provide explicit solutions to all 5d fields, including all vector multiplets, up to a single function obeying the sourced Liouville equation.

Having finished the discussion of a single Killing vector we move on to discuss the case of two commuting Killing vectors in section \ref{toricsection},  focusing on the situation where the Killing vectors are Hamiltonian and the geometry of the base is toric. Analyzing the additional Killing vector along the lines of section \ref{kvclass} one finds that one of the two Killing vectors will always leave the axidilaton $\tau$ invariant (which explains our focus on that case in section \ref{invKV}) so that the pair of Killing vectors can be of type $II$, $IP$, $IH$ or $IE$.  Furthermore, the space-time dependence of the scalar $\tau$ is completely fixed, up to some integration constants and a discrete choice related to the type of symmetries. In particular, in class $II$ the axidilaton is forced to be constant, and as a warm-up we first review this case in our formalism. We focus on the case where one of the Killing vectors is translational and the other is rotational and show how
the solutions to the Toda equation as well to the equations governing all 5D fields, are reduced to specifying a number of axially symmetric harmonic functions in $\RR^3$. We then turn to the situation with nonconstant dilaton. In order  to have a dynamical axidilaton in the presence of two commuting symmetries, one is forced to have $\tau$ transform with a non-trivial U-duality under one of the symmetries.
 Furthermore if that symmetry is along a compact direction it means $\tau$ will have monodromy, which signals the presence of brane sources.   In section \ref{intrM2} we discuss in  more detail how and under which conditions the solutions with toric symmetry can be interpreted as backreacted M2 branes that are extended in the 5d external directions.  Finally, in section \ref{sectoricToda}, we analyze the special class of separable solutions to the Toda equation under the additional assumption of toric symmetry. This concludes our general analysis of dynamic axidilaton solutions.

After our rather abstract and technical discussion in the first sections we illustrate all the features discussed  there in a number of examples in section \ref{secexs}. We start by reviewing some physically interesting  solutions with a toric hyperk\"ahler base in our formalism,
including cases where the base is `ambipolar' and changes signature in some region, commenting  on the generalization of   toric
geometry which governs these spaces.
We then turn to solutions with axidilaton, focussing on those solutions which describe backreacted M2-branes placed in a background with a toric base. We discuss in detail backreacted branes in flat space and the highly symmetric G\"odel$\times$S$^2$ solution \cite{Levi:2009az} which, as we will argue, arises from a distribution of branes in the AdS$_3\times$S$^2$ background. We also comment on the solutions describing individual branes in the Eguchi-Hanson,  AdS$_3\times$S$^2$, and AdS$_2\times$S$^3$ backgrounds,  the latter being of interest for  the black hole deconstruction proposal
\cite{Denef:2007yt}. These solutions are   fully specified by a single function satisfying an ordinary non-linear differential equation.
They will be discussed in more detail using the tools developed in this paper in a  future publication \cite{JRDVDB}.

\section{Characterizing supersymmetric 5D axidilaton solutions}\label{charsec}
In this section we first review some basics of 5D  N=1 supergravity and its relation to M-theory compactified on a Calabi-Yau manifold. We point out some of the essential equations and geometric structure that govern general supersymmetry preserving solutions to this theory, following the work of \cite{Bellorin:2006yr}. We analyze this general structure in more depth in the case of truncation to the axion-dilaton scalars inside the universal hypermultiplet and show how solutions are completely determined by a choice of holomorpic axion-dilaton profile and a single remaining complex non-linear equation, essentially a sourced Monge-Amp\`ere equation.

\subsection{N=1 supergravity from M-theory on a Calabi-Yau}
Local supersymmetry in 4+1 dimensions requires a minimum of 8 real supercharges, which we will call 5D N=1 supersymmetry. In this work we will consider ungauged N=1 supergravity theories that apart from the gravity multiplet contain couplings to vector multiplets and  hypermultiplets. Let us briefly review the bosonic field  content and the geometry governing such a theory. The bosonic fields of  the gravity multiplet are the metric and the graviphoton $A^0$.  Each of the $n_v$  vector multiplets contains a massless vector $A^x$ and a real scalar $\f^x,\ x = 1 , \ldots , n_v$. The vector multiplet sector is governed by very special real geometry \cite{Gunaydin:1983bi, deWit:1992cr}. The matter vectors  $A^x$  can be combined  with $A^0$ into a column  vector $A^I, I = 0, \ldots , n_V$ transforming as a vector under an $SO(n_v +1)$ global symmetry of the theory.
Similarly it is convenient to describe the scalar manifold in terms of  $n_v +1$ homogeneous coordinates $Y^I(\f)$ satisfying a constraint
 \be
 D_{IJK} Y^I Y^J Y^K = 6.
 \ee
Here, $ D_{IJK}$ is a totally symmetric $SO(n_v +1)$ tensor which completely determines the metric $g_{xy}(\f)$ on the scalar manifold and the scalar-dependent  kinetic term $a_{IJ} (\f)$ for the vectors. For explicit expressions we refer to  \cite{Bellorin:2006yr}, appendix A.3\footnote{We follow essentially the conventions of \cite{Bellorin:2006yr}, with exception of the metric  signature (ours is
mostly plus), and the quantities $h^I$ and $C_{IJK}$ in \cite{Bellorin:2006yr} are related to ours as $h^I = Y^I/\sqrt{3},\ C_{IJK} = \sqrt{3}D_{IJK}/2$}. Each of the $n_h$ hypermultiplets contains 4 real scalars which we collectively denote as $q^X, \ X = 1, \ldots , 4 n_h$, whose target space is a quaternionic K\"ahler manifold  \cite{Bagger:1983tt} with metric $g_{XY}(q)$.

The bosonic part of the most general 2-derivative  supersymmetric Lagrangian describing these fields
is
\bea
S &=& \int d^5 x \sqrt{-g} \left[R + \half g_{xy}(\f) \pa_\m \f^x \pa^\m \f^y + \half g_{XY} (q) \pa_\m q^X \pa^\m q^Y - {1 \over 4} a_{IJ} (\f)F^I_{\m\n}  F^{J\m\n}\right]\nonumber\\
&& + {D_{IJK}\over 6} \int F^I \wedge F^J \wedge A^K.\label{sugraS}
\eea

We will be especially interested in the 4+1 dimensional theory arising from compactifying 11-dimensional supergravity on a Calabi-Yau manifold $X$
\cite{Cadavid:1995bk}. The field content of the 5D theory is now directly related to the Hodge-numbers $h_{(i,j)}$ of $X$. Besides the gravity multiplet, this theory contains $h_{(1,1)}-1$ vector multiplets where the tensor $ D_{IJK}$ determining the real geometry is  given by the intersection matrix on $X$. The hypermultiplets consist of  the universal hypermultiplet \cite{Cecotti:1988qn}, whose  couplings are independent of the topology of $X$, and an additional $h_{(2,1)}$ hypermultiplets. In this work we will consider solutions where only the universal hypermultiplet plays a role.
Its bosonic fields, viewed as 2 complex scalars, arise as follows. The first one of these   is an axion-dilaton-like  field, with a real part which is the Hodge dual of the three form with legs in the 5D spacetime, and
   an imaginary part coming from the volume modulus of $X$.  We will refer to this field as the axidilaton $\t$. The other complex scalar arises from the three form  modes proportional to the $(3,0)$ and $(0,3)$ form on $X$. The hypermultiplet  moduli space is a direct product of the universal hypermultiplet moduli space and that of the remaining $h_{(2,1)}$ hypermultiplets. The universal part of the hypermultiplet moduli space  is the homogeneous quaternionic space
${SU(1,2) \over U(2)}$.

\subsection{Structure of supersymmetric solutions}
In \cite{Bellorin:2006yr}, the general structure of supersymmetric solutions of 4+1 dimensional supergravity with vector- and hypermultiplets was analyzed, extending the pioneering work  on minimal supergravity in \cite{Gauntlett:2002nw}. The idea is to assume the existence of a Killing spinor and analyze how the Killing spinor equations constrain the bosonic fields constructed out of Killing spinor bilinears. We now  briefly review the results of this analysis. A first spinor bilinear yields a Killing vector, which in the current work will be assumed to be everywhere timelike. Choosing an adapted coordinate, the metric is of the form
\be
ds^2 = - f^2 ( dt + \xi)^2 + f^{-1} ds^2_4 \label{5Dmetric}
\ee
where $ds^2_4$ denotes the  Euclidean metric on a 4-dimensional base manifold which we will refer to as the {\em base}.

Let's first discuss the BPS equations which constrain the geometry of the base, which do not involve the vector multiplets. There exist three
 selfdual 2-forms  $\F^a,\ a= 1,2,3$  which endow the base with an  almost quaternionic structure:
\bea
\F^{a} &=& \star_4 \F^a\\
\F^{a A}_{\ \  C} \F^{b C}_{\ \  B} &=& - \d^{ab}\d^A_B + \e^{ab}_{\ \ c} \F^{c A}_{\ \ B}. \label{quatern}
\eea
Our index convention is as follows:  $A,B= 1, \ldots, 4$ are 4D tangent space indices, while 4D curved indices will be denoted by $\m,\n = 1, \ldots, 4$.
Note that these relations are invariant under local $SO(3)$ transformations under which the $\F^a$ transform as a triplet.
The BPS equations governing these two-forms are
\be
\nabla_\m \F^{a}_{\ BC} +  \e^{a}_{\ bc} A^b_\m \F^{c}_{\ BC}=0\label{eq1}.
\ee
 With a slight abuse of notation we have denoted by $A^a \equiv A^a_X d q^X$  the pullback  of the $SU(2) \subset SO(4)$ part of the spin connection on the quaternionic hypermanifold.
 We note that, when the hyperscalars are constant, this equation tells us that the $\F^a$ must be covariantly constant and hence endow the   base with a hyperk\"ahler structure.

The hypermultiplet scalars $q^X$ parameterize a map from the  base into the quaternionic target space which is constrained by the BPS condition
\be
(d q^X)_A = \F^{a\ B}_{\ A} (d q^Y)_B  J^{a\ X}_{\ Y}\label{eq2}
\ee
 where the $J^{a\ X}_{\ Y}$ form the quaternionic structure of the hypermultiplet target space.
This type of map was called a {\em quaternionic map} in \cite{Bellorin:2006yr}. One of the purposes of this work is to demystify this condition in the simplest context when only the axidilaton is turned on, where we will see that it reduces to a simple holomorphicity condition.

In addition to equations (\ref{eq1}, \ref{eq2}), there are  additional BPS conditions which  determine the warp factor  $f$ and the one-form $\xi$ in (\ref{5Dmetric}), as well as the the vector multiplet scalars $Y^I$ and the Maxwell field strengths $F^I$. Supersymmetry relates all of these fields  to $n_V+1$  harmonic anti-selfdual 2-forms $\Theta^I$ on the 4D base as follows:
\bea
d \Theta^I &=& 0, \qquad  \star_4 \Theta^I= -\Theta^I\label{VMeq1}\\
\nabla^2_4 (f^{-1} Y_I) &=& {1 \over 2} D_{IJK} \Theta^J \cdot  \Theta^K\label{VMeq2}\\
d \xi - \star_4 d\xi &=&   \half f^{-1} Y_I \Theta^I\label{VMeq3}\\
F^I &=& - d( f Y^I ( dt + \xi )) + \Theta^I\label{fluxeqs}
\eea
where $Y_I \equiv D_{IJK} Y^J Y^K,\ \a \cdot  \b= \a_{\m\n} \b^{\m\n}$ and the vector multiplet scalars $Y^I$  satsify the constraint
$D_{IJK} Y^I Y^J Y^K = 6$.

\subsection{Axidilaton solutions}
In the 5D supergravity theory arising from 11-dimensional supergravity on a Calabi-Yau manifold,  the hypermultiplet moduli space is a direct product of a universal hypermultiplet
component and a component associated to the remaining $h_{(2,1)}$ hypermultiplets. The theory therefore allows a consistent truncation to the class of solutions where only the universal hypermultiplet is turned on. In that case the 4-dimensional hypermultiplet moduli space is
${SU(1,2) \over U(2)}$, see appendix \ref{hypermod} for a brief review and conventions. The moduli space has an $SU(1,2)$ isometry which acts as a U-duality group of the 5D fields.
Note that the two-forms $\F^a$ also transform under U-duality.
Indeed, from (\ref{eq1}) we see that  the $\F^a$ not only transform as two-forms under diffeomorphisms, but also rotate into each other under local frame rotations of the hypermultiplet target
space. This determines  how the $\F^a$  transform under U-duality: an $SU(1,2)$ U-duality induces an
 $SO(4) \simeq SU(2)\times SU(2)'$ frame rotation in target space, and the $\F^a$ rotate into each other under the pullback of the $SU(2)$ part. The metric
 on the 4D base is however U-duality invariant.

Now  we consider hypermultiplet solutions where only the axidilaton is turned on. This means we look at a further  consistent truncation of the theory where two of
 the scalars $q_3, \ q_4$ are  constant while $q_1, \ q_2$ can fluctuate. Without loss of generality, we will set $q_3=q_4=0$ in what follows. We use the standard notation $\t = \t_1 + i \t_2$ for the axidilaton, with
\be
q^1 = - \t_2, \qquad q^2 = -\t_1.
\ee
The hyperscalar  metric (\ref{hypermetric}) on this submanifold is
\be
ds^2 = {d\t d\bar \t \over 4 \t_2^2}.
\ee
The part of the original U-duality group which leaves the subspace $q_3= q_4=0$ invariant is $SU(1,1)\simeq SL(2,\RR)$, which acts on $\t$ as the familiar fractional linear transformations.

The $SU(2)$ connection $A^a$, which in our conventions is given by (\ref{su2conn}), becomes
\bea
A^1 &=&  A^2 =0\\
A^3 &=& -{ d\t_1 \over 2 \t_2}.
\eea
For pure axidilaton solutions eqs. (\ref{eq1}) simplify to, defining $\F^\pm = \F^1 \pm i \F^2$:
\bea
\nabla_\m \F^\pm \pm i A^3_\m \F^\pm &=&0\label{ad1}\\
\nabla_\m \F^3 &=& 0
 \eea
The last equation states that the almost complex structure $\F^3$ is covariantly constant.
Hence when turning on only the axidilaton, the base is still K\"ahler, with K\"ahler form $\F^3$, but it will  in general no longer be hyperk\"ahler. Note that the completely antisymmetric part of (\ref{ad1}) can be written as
\be
d \F^\pm \pm i A^3 \wedge \F^\pm =0\label{ad1p}.
\ee
As discussed above, the  forms $\F^a$ transform  under U-dualities. Since $A^3$ transforms under fractional linear transformations as
\begin{equation}
\tau\rightarrow \frac{a\tau+b}{c\tau+d}\qquad\mbox{and}\qquad A_3\rightarrow A_3-d \rm{Im} \log(c\tau+d)
\end{equation}
the two-forms $\F^\pm$ must transform by a phase in order for (\ref{ad1}) to remain invariant:
\be
\F^\pm \to e^{\pm i \rm{Im} \log(c\tau+d) } \F^\pm. \label{Fpmmodtransf}
\ee

Exploiting the fact that the metric is K\"ahler with respect to $\F^3$, we introduce adapted  complex coordinates $w^1, w^2$
as well as a unitary frame  $\vf^1, \vf^2$ of $(1,0)$ forms such that
\bea
ds^2 &=&  g_{i\bar j} dw^i d \bar w^j= \vf^1  \bar \vf^1 + \vf^2  \bar \vf^2 \\
\F^3 &=& { i\over 2} g_{i\bar j} dw^i  \wedge d \bar w^j  = - {i } \left( \vf^1 \wedge \bar \vf^1 + \vf^2 \wedge \bar \vf^2 \right).\label{vielb}
\eea
It follows from (\ref{quatern}) that $\F^+$ and $\F^-$ are of type $(2,0)$ and $(0,2)$ respectively, and that
$\F^+ \wedge \F^- = 2 \F^3 \wedge \F^3$. Using also that $\F^- = \overline{\F^+}$ fixes $\F^+$ up to a real phase  $\l$
\be
\F^+ =  e^{i\l} \vf^1 \wedge \vf^2.\label{Phiplframe}
\ee
(The phase $\l$ could be absorbed by a frame rotation, but we prefer to keep our frame arbitrary.)

First, let's  analyze the equations (\ref{eq2})  for the axidilaton in this frame. Using (\ref{quatJs}) one finds that they are equivalent to
\be
(d\t)_{ \bar \vf^1} =(d \t)_{ \bar \vf^2}=0
\ee
where the subscripts denote components in the unitary basis (\ref{vielb}). In other words,  the quaternionic map condition (\ref{eq2}) here
simply states that $\t$ must be a holomorphic function:
\be
\pa_{\bar w^1} \t = \pa_{\bar w^2}\t =0.
\ee

Now we turn to the antisymmetrized equations (\ref{ad1p}) for $\F^\pm$ which reduce to
\be
d\F^+ - i {d\t_1 \over 2 \t_2} \wedge \F^+ =0\label{Fpmeqs}
\ee
and the complex conjugate thereof.
Arguments similar to the one above (\ref{Phiplframe}) show that (\ref{quatern}) determines $\F^+$  in the coordinate basis up to a real phase $\a$:
\be
\F^+ =  \sqrt{g_\CC}  e^{i\a} dw^1 \wedge dw^2.\label{Phiplcoord}
\ee
 where $g_\CC\equiv \det \{g_{i\bar j} \} = \sqrt{\det g}$.
Using that $d=\partial+\bar\partial$ and the holomorphicity  of $\tau$ one finds that (\ref{Fpmeqs}) is equivalent to
\begin{equation}
\bar\partial \log {g_\CC e^{2 i \a} \over \t_2}=0
\end{equation}
This implies that there exists a holomorphic function $h(w^1,w^2)$ such that
\be
{g_\CC e^{2 i \a} \over \t_2} = e^h\\
\ee
Furthermore, since both $g_{\mathbb{C}}$ and $\tau_2$ are strictly positive and $\alpha$ is real it follows, setting $h \equiv h_1 + i h_2$,
that
\be
h_2 = 2 \a\label{h2alpha}
\ee
Hence we obtain the following constraint on the base metric:
\be
g_\CC =  \t_2 e^{h_1}.\label{geq}
\ee

In summary, a general supersymmetric solution is specified by two holomorphic functions $\t, h$ and a metric which is K\"ahler and satisfies (\ref{geq}).
The latter two conditions can be combined into a nonlinear   differential equation for the K\"ahler  potential, which for later convenience
we normalize as $g_{i \bar j} = 4 \calk_{i \bar j}$,
\be
\calk_{1 \bar 1} \calk_{2 \bar 2} - \calk_{1 \bar 2} \calk_{2 \bar 1}  = {\t_2 e^{h_1}\over 16}.\label{defMA}
\ee
This is a nonlinear partial differential equation of the  Monge-Amp\`ere type, see e.g. \cite{Pogorelov}.
In the case of constant axidilaton, it is the familiar Monge-Amp\`ere equation expressing that the K\"ahler base is Ricci flat,  which  is equivalent to the hyperk\"{a}hler condition in 4 real dimensions. A non-constant axidilaton backreacts on the metric  by introducing  a source in the RHS of the  equation and deforming the geometry away from being hyperk\"ahler.
The Ricci tensor of the base  is
\be
R_{i \bar j} = - i \pa_i \pa_{\bar j} \ln g_\CC =  - i \pa_i \pa_{\bar j} \ln \t_2
\ee
and the two-forms $\F^+$ are given by
\be
\F^+ ={\sqrt{ \t_2 }  e^{h \over 2}\over 4} dw^1 \wedge dw^2, \qquad \F^- = \overline{\F^+}\label{Fplsol}
\ee
Although so far we have only imposed the fully  antisymmetric part (\ref{ad1p}) of the equations (\ref{ad1}), we have checked  that the remaining equations in (\ref{ad1}) are automatically satisfied.

So far we have not yet chosen specific holomorphic coordinates and we are free to make  holomorphic coordinate transformations. We should note however that, while $\t$ and $\calk$ transform as scalars, $h$ must transform  nontrivially such that eq. (\ref{defMA}) is invariant.  In particular, $e^h$ must be a density of weight 2 so that under $w \to \tilde w(w)$  the field $h$ transforms as
\be
\tilde h(\tilde w) = h(\tilde w) + 2 \ln \det \left({\pa \tilde w^i \over \pa  w^j}\right)\label{htransf}
\ee
The infinitesimal transformation of $h$ generated by a holomorphic vector field $k$ is by  definition the Lie derivative, which has an extra term compared to the Lie derivative of a scalar field:
\be
\d_k h \equiv \call_k h  =k^i h_{,i} + 2 \pa_i k^i\label{Lieders}
\ee
Note that in principle this implies one can always (locally) choose coordinates in which $h$ becomes trivial. But as will become clear in the following such coordinate choice makes other aspects of the solutions less transparent and so we prefer to keep $h$ free and preserve manifest holomorphic coordinate invariance for now.

\subsection{Redundancies}\label{reds}
Recapitulating, we have described the configuration space of supersymmetric axidilaton solutions  in terms of two holomorphic functions $\t$ and $h$ (recall that the imaginary part of $h$ is the phase of $\F^+$) and a real function $\calk$ satisfying eq. (\ref{defMA}). Our description is however redundant since the following symmetry transformations on $\t, h , \calk$ produce equivalent configurations:
\begin{itemize}
\item {\bf U-duality transformations.} The  $SL(2,\RR)$ U-duality transformations act as fractional linear transformations on $\t$. Since the two-forms
$\F^\pm$ are charged under U-duality and transform as  (\ref{Fpmmodtransf}), it follows  from (\ref{h2alpha}) that  $e^h$ also has a nontrivial transformation law and is in fact a modular form of weight 2:
\bea
\tilde \t &=& { a\t + b \over c\t + d}; \qquad ad-bc =1\\
\tilde h &=& h+ 2 \log(c\tau+d) \\
\tilde \calk &=& \calk. \label{Uduality}
\eea
Note that this implies that also (\ref{defMA}) is invariant.
Let us also write down the infinitesimal version of this transformation law. Parameterizing a general $sl(2, \RR)$ Lie algebra element as
\be
Q = r L_0 + q L_1 + p L_{-1} = \left(\begin{array}{cc} {r\over 2} &p\\-q& -{r\over 2}\end{array}\right)
\ee
for  $p,q,r \in \RR$, we have
\bea
\d_U \t &=& p + r \t + q \t^2\\
\d_U h &=& - r - 2 q \t\\
\d_U \calk &=& 0
\eea
\item {\bf K\"ahler transformations.} As always, the K\"ahler potential is only defined up to addition of the real part of a holomorphic function:
\be
\d_K \calk = \e (w) + \bar \e (\bar w ),\qquad \d_K \t= \d_K h = 0 \label{Kahlertransfo}
\ee
\item {\bf Global U(1) rotations of $\F^\pm$.} Finally, we are free to rotate $\F^\pm$ by a constant   phase $e^{\pm i s}$, which corresponds to an imaginary shift of $h$:
\be
\d_{rot} h= i s, \qquad \d_{rot} \t = \d_{rot} \calk =0\label{globalU1}
\ee
This corresponds to rotating the Killing spinor by an overall phase.
\end{itemize}
Hence it's not quite correct to think of  $\t, h, \calk$ as functions, rather they are sections of appropriate line bundles that can undergo transformations of the above types when going to a different coordinate patch.

In particular, when going around a closed curve, the $\t$ and $h$ fields can pick up a monodromy by a U-duality transformation, which signals
a degeneration of the internal Calabi-Yau manifold. Recalling that $\t_1$ is the Hodge dual of the M-theory three-form with legs in the 5D noncompact space, it is easy to see that a monodromy  $\t \to \t + 1$ signals the presence of an M2-brane extended in the 5D noncompact space and smeared over the internal  Calabi-Yau.
More general $SL(2,\ZZ)$-valued monodromies signal the presence of exotic branes which do not descend from 11D M-branes  \cite{deBoer:2010ud},\cite{deBoer:2012ma}. Note that in the present case the exotic branes are geometric from the 5D point of view, the 5D metric being single-valued when encircling these objects.

\section{Structure of solutions with extra Killing vectors}\label{kilvsec}
In this section we discuss the simplifications which occur in the presence of one or two additional Killing vectors which preserve the  structure imposed by supersymmetry.
 In the case of two Killing vectors we will focus on the situation where the base has  (generalized) toric geometry; the geometry of the   base is then fully specified by a single function satisfying an ordinary non-linear differential equation.
\subsection{Single compatible Killing vector: classification}\label{kvclass}
We will from now on focus on  supersymmetric solutions which admit, besides the timelike Killing vector constructed out of the Killing spinor itself,  an additional Killing vector on the  base. It is natural to restrict attention to   Killing vectors which   not only preserve  the metric but also the additional  structure imposed by supersymmetry discussed  in the previous section. We will call such Killing vectors {\em compatible} with the supersymmetric structure.
For example, we will consider only Killing vectors which preserve the complex structure with K\"ahler form $\F^3$, and hence  will
 restrict our attention to holomorphic Killing vectors. We want the Killing vector to furthermore preserve $\t, h $ and $\calk$, which is certainly the case if these functions are strictly invariant, i.e. their appropriately defined Lie derivatives (see e.g. (\ref{Lieders})) vanish. This is however too strong a requirement in view of the redundancies discussed  in paragraph \ref{reds} above:
 it is sufficient if their holomorphic transformation  can be compensated for by a combination of a U-duality (\ref{Uduality}), a K\"ahler transformation (\ref{Kahlertransfo}) and a U(1) rotation (\ref{globalU1}). In other words, we must have a transformation law of the form
\be
\call_k F = \d_U F + \d_K F + \d_{rot} F\label{geninv}
\ee
where $F $ stands for any of the fields $\t,h,\calk$.
We will now explore how this requirement constrains the fields. To simplify the discussion we choose local holomorphic coordinates $w^1, w^2$ such that $w^1 = x^1 + i \theta^1$ is adapted to the Killing vector  $k$:
\be k= \pa_{\theta^1} = i (\pa_{w^1}-\pa_{\bar w^1}).\ee

Starting with the axidilaton $\t$, the holomorphic reparametrization must induce an infinitesimal fractional  linear transformation, $\call_k \t = \d_U \t$, so that $\t$ must be a solution of
\be
 i \pa_{w^1} \t =  p + r \t + q \t^2 \label{tauinv}
\ee
for some $p,q,r$. The  general solution of this equation has four subcases:
\be
\t  = \begin{cases} \tilde \t (w^2) & {\rm for\ }  Q= 0 \\
-i p w^1+ \tilde \t (w^2)& {\rm for\ } r=q= 0, p\neq 0\\
 - {p \over r} +  \tilde \t (w^2) e^{i r w^1} & {\rm for\ } q= 0, r\neq 0\\
 -{r \over 2 q} - i {\sqrt{\det Q }\over q} \tanh\left( \sqrt{\det Q } w^1 + i \tilde \t (w^2)\right)&   {\rm for\ } q\neq 0, \det Q\neq 0\end{cases}\label{taugensol}
\ee
with $\tilde \t$ an arbitrary function of $w^2$.

We can simplify these expressions a bit by choosing a convenient U-duality frame: under a change of U-duality frame, the $sl(2,\RR)$ element $Q$ is conjugated by an $SL(2,\RR)$ group element, which we can use to pick a simple representative within the same conjugacy class. Conjugacy classes are labeled by the value of $\det Q = pq - r^4/4$. There are four distinct cases depending  on whether $\t$ is invariant ($I$) or transforms by element of an elliptic ($E$), hyperbolic ($H$)  or parabolic ($P$)  conjugacy class. The representative we will choose in each of these classes is shown in table \ref{table1}. Note that if $k=\pa_{\theta^1}$ generates a compact U(1) isometry, in all except the invariant cases $\t$ picks up a monodromy when circling around the U(1) direction, which  signals the presence of M2 (in the parabolic case) or exotic (in the hyperbolic and elliptic cases) brane charge.
\begin{center}\begin{table}\begin{tabular}{c|c|c|c}
Class & representative & $e^Q$ & $\t$\\ \hline
I: invariant, $Q =0$ & $p=q=r=0$ & $\left(\begin{array}{cc} 1 &0\\0&1\end{array}\right)$ & $\tilde \t (w^2)$ \\  \hline
P: parabolic, $\det Q =0$ & $r=q=0$& $\left(\begin{array}{cc} 1 &p\\0&1\end{array}\right)$ &$-i p w^1+ \tilde \t (w^2) $\\ \hline
H: hyperbolic, $\det Q <0$  & $p=q=0$&$\left(\begin{array}{cc} e^{r\over 2} &0\\0&e^{-{r\over 2}}\end{array}\right)$ & $\tilde \t (w^2) e^{i r w^1}$ \\ \hline
E: elliptic, $\det Q > 0$ & $r=0, p=q $ &$\left(\begin{array}{cc} \cos q &\sin q\\-\sin q&\cos q \end{array}\right)$ &$- i \tanh (q  w^1 + i \tilde \t (w^2))$
\end{tabular}
\caption{Convenient choices of representative within each U-duality conjugacy class.}\label{table1}
\end{table}
\end{center}

Turning next to  the field $h$, the invariance condition (\ref{geninv}) says that its Lie derivative amounts to a combined U-duality and U(1) rotation,  $\call_k h = \d_U h + \d_{rot} h$, leading to
\be
 i \pa_{w^1} h = - r - 2 q \t + i s_1 \label{hinv}
 \ee
If $Q \neq 0$ we have (except possibly at isolated points)  $\pa_{w^1} \t  \neq 0$ and (\ref{hinv}) can be rewritten using (\ref{tauinv}) as
$\pa_{w^1} (h - s_1 w^1) = \pa_{w^1} ( - \ln i \pa_{w^1} \t )$. Hence (\ref{hinv}) integrates to  the simple general solution
\be
h = \begin{cases}  - \ln ( i \pa_{w^1} \t )+ s_1 w^1 + \tilde h (w^2)& {\rm for\ } Q\neq 0 \\
  s_1 w^1 + \tilde h (w^2) & {\rm for\ } Q= 0\end{cases} \label{h1KV}
\ee
with $\tilde h$ another arbitrary function of $w^2$.
 In table \ref{table2} we list the expressions for $h$ for the four types of Killing vector as well as for the combination $\t_2 e^{h_1}$ entering in the equation (\ref{defMA}) for the K\"ahler potential.

Let us also introduce some commonly used terminology related to
 the U(1) term in the transformation of $h$ with parameter  $s_1$. When $s_1$ is zero, the two-forms $\F^\pm$ are invariant under the isometry (see (\ref{Fplsol})) and the isometry is usually called {\em translational} following \cite{Boyer:1982mm}.
When $s_1$ is nonzero, $\F^\pm$ have charge $\pm s_1$ and the isometry was called {\em rotational} in  \cite{Boyer:1982mm}.  Note that in this case we have a compact $U(1)$ isometry.
  If we normalize $\theta^2$ to have period $2 \p$, we see from (\ref{Fplsol}) that requiring $\F^\pm$ to be single-valued when going around the  $U(1)$  direction requires that $s_1$ is quantized in units of 2. Since $\F^\pm$ is constructed out of a spinor bilinear we see that  when $s_1/2$ is even, (resp. odd), the Killing spinor has even (resp. odd) spin structure around the  $U(1)$  direction.
 
\begin{center}\begin{table}\begin{tabular}{c|c|c}
Class & $h$ & $ \t_2 e^{h_1}$ \\ \hline
I &  $ h = s_1 w^1 + \tilde h(w^2)$ & $ \tilde \t_2  e^{\tilde h_1 + s_1 x^1} $ \\ \hline
P &$ h = s_1 w^1 + \tilde h(w^2)$ & $ (\tilde \t_2- p x^1) e^{\tilde h_1+ s_1 x^1} $\\\hline
H& $ h =  ( i r+ s_1) w^1 +\tilde h(w^2)$&$(\tilde \t_2 \cos r x^1 - \tilde \t_1 \sin r x^1) e^{\tilde h_1+ s_1 x^1}$ \\\hline
E& $ h = 2 \ln \cosh ( q w^1 + i \tilde \t(w^2) + s_1 w^1 + \tilde h(w^2))$  & $\half \sinh (2 \tilde \t_2- 2 q x^1) e^{\tilde h_1+ s_1 x^1} $ \\ \hline
\end{tabular}
\caption{The functions $h$ and $\t_2 e^{h_1}$ for the various types of Killing vector.}\label{table2}
\end{table}
\end{center}

Now let's turn to the K\"ahler potential $\calk$. In principle we can only ask that $\calk$ is invariant up to a K\"ahler transformation, that is,
\be
\pa_{\theta^1} \calk = f(w^i) + \bar f (\bar w^i).
\ee
It is easy to see however that, by making a suitable K\"ahler transformation, we can make the K\"ahler potential locally invariant\footnote{Obstructions can arise when trying to do this simultaneously for several noncommuting Killing vectors.}:
\be
\pa_{\theta^1} \calk = 0.
\ee
Choosing  an invariant representative for the K\"ahler potential is consistent with the Monge-Amp\`ere equation (\ref{defMA}) for $\calk$  since one can show that, as a consequence of (\ref{tauinv},\ref{hinv}),  $\t_2 e^{h_1}$ is independent of $\theta^2$, as can be seen in the explicit solutions above. The Monge-Amp\`ere equation (\ref{defMA}) then reduces to
\be
\calk_{w^2 \bar w^2}\calk_{x^1x^1}  - \calk_{w^2 x^1 }\calk_{\bar w^2 x^1 } = { \t_2 e^{h_1} \over 4}.\label{MAKV}
\ee
Incidentally, it follows for this choice of K\"ahler potential representative that
\be
i_{\pa_{\theta^1}} \F_3 = - 2 (\bar \pa \calk_{w^1} + \pa \calk_{\bar w^1})= - d  (\calk_{w^1} + \calk_{\bar w^1})=- d\calk_{x^1}\label{killingpot}
\ee
Hence, if $\calk_{x^1}$  extends to a globally well-defined function, the Killing vector is Hamiltonian  with moment map $\calk_{x^1}$.

\subsection{Simplifications when $\t$ is invariant}\label{invKV}
From now on we will focus on solutions which have an adapted (in the sense explained in section \ref{kvclass}) Killing vector
under which $\t$ is invariant, in other words they belong  to the class $I$ of the classification above. This doesn't mean that our analysis of the
 other classes (which we labeled by $P$, $H$ and $E$) of Killing vectors was in vain  however, since in section \ref{toricsection} we will consider solutions with  a second Killing vector under which  $\t$ need not  be invariant. Anticipating this we will slightly change notation and switch the roles of $w^1$ and $w^2$ with respect to the previous section, i.e. the  invariant Killing vector is $k= \pa_{\theta^2} = i (\pa_{w^2}-\pa_{\bar w^2}),$ and $\t$ and $h$ are of the form
 \bea
 \t &=& \tilde \t (w^1)\\
 h &=& \tilde h (w^1) + s_2 w^2.
 \eea

\subsubsection{The Toda frame}\label{Tframe}
In the case of invariant $\tau$ we just mentioned, the equation (\ref{MAKV}) determining the 4D metric can be cast in a more manageable form by making a Legendre transformation.
We define a potential $\calv$ which is the  Legendre transform of $\calk$ with respect to $x^2$:
\be
\calv (w^1, \bar w^1, y_2 ) = x^2 y_2 - \calk; \qquad y_2 = \calk_{x^2}
\ee
It's useful to introduce a new special symbol $\Psi$ for the $y_2$ derivative of $\calv$:
\be\label{Todapot}
\Psi (w^1, \bar w^1, y_2 )  \equiv \calv_{y_2} (w^1, \bar w^1, y_2 )   = x^2(w^1, \bar w^1, y_2 ).
\ee
 We will refer to $\Psi$ as the {\em Toda potential} for reasons we will now explain.
Using the Legendre transformation of second derivatives
 \bea
 \calk_{x^2 x^2} &=& {1 \over \calv_{y_2 y_2}}\\
 \calk_{w^1 x^2} &=& - {\calv_{w^1 y_2}\over \calv_{y_2 y_2}}\\
 \calk_{w^1 \bar w^1}&=& {\calv_{w^1 y_2}\calv_{\bar w^1 y_2} \over \calv_{y_2 y_2}}- \calv_{w^1 \bar w^1}.
 \eea
we see that the Monge-Amp\`ere equation (\ref{MAKV}) reduces to the following equation for $\Psi$,
\be
4 \Psi_{w^1\bar w^1} + {\tilde  \t_2 e^{\tilde h_1} \over s_2 }  (e^{s_2 \Psi})_{y_2y_2} =0 \label{Toda}
\ee
When the axidilaton is constant this equation is reduces, after a holomorphic reparameterization setting $\h$ to zero,  to  the $SU(\infty )$ Toda equation or, in the limit $s_2\rightarrow 0$, to the 3D flat Laplace equation. These are the well-known equations describing hyperk\"ahler metrics with a rotational/translational isometry \cite{Boyer:1982mm, Gegenberg}. See \cite{Saveliev,Savelev:1993if} for more information on  the $SU(\infty )$ Toda equation and its solutions.

Defining also
\bea
K^0 &\equiv& \Psi_{y_2}\label{defKu0}\\
\chi &\equiv&  - 2 \Im (\Psi_{w^1} dw^1)\label{defom0}
\eea
the base metric and K\"{a}hler form are
\bea
ds^2_4 &=&  K^0 ds^2_3 + {1 \over K^0} \left( d \theta^2 +
\chi \right)^2 \label{4dmetrrot}\\
ds^2_3 &=& {dy_2^2 } + \tilde \t_2 e^{\tilde h_1}  e^{s_2  \Psi} dw^1 d\bar w^1\label{3dmetrrot}\\
\F^3 &=& {i \over 2} K^0 \tilde \t_2 e^{\tilde h_1+ s_2 \Psi} dw^1 \wedge d\bar w^1 + dy_2\wedge \left(d\theta^2 - 2 {\rm Im} (\Psi_{w^1} dw^1) \right)\label{Kformrot}
\eea
 Note that (\ref{Toda}) can be formally written as a Laplace equation $\D \Psi =0$ with respect to the  3D metric (\ref{3dmetrrot}), with the proviso that  the 3D metric itself depends on $\Psi$ when $s_2 $ is nonzero. The  3D metric is is not flat in general and its scalar curvature is given by
 \be
 R^{(3)} = {1\over  \tilde \t_2 e^{\tilde h_1+ s_2 \Psi}}\left( {s_2 \over 2} (s_2 (K^0)^2 + 2 K^0_{y_2}) + {\tilde \t' \bar{\tilde \t}'\over \tilde \t_2^2}\right).\label{Ricci3d}
\ee

\subsubsection{Structure of 5D multibrane solutions}\label{mbsection}
Having determined the geometry of the base in the presence of a Killing vector, we now turn to the solution of the  equations (\ref{fluxeqs}) which determine the full 5D metric  as well the bosonic fields in  the vector multiplets.
In the absence of hypermultiplets the analysis of the system of equations (\ref{fluxeqs}) on a base with rotational isometry   was performed in \cite{Bena:2007ju} and is easily generalized to include the axidilaton. We will also expand on the  discussion
given there, most notably in the discussion around (\ref{spectrflow}),(\ref{multicentconstr}).
We choose to write the solution in a form which allows easy comparison
with the more extensively  studied solutions with translational Killing vector, to which they should reduce when the parameter $s_2$ is taken to zero. More details can be found  in Appendix \ref{phase2}.

Starting from a solution (\ref{Toda}, \ref{4dmetrrot}, \ref{3dmetrrot}) for the 4D base manifold, the general 5D solution depends  on an additional set of functions
 $K^I,K_I,K _0$ and a one-form $\o$.
Let us first discuss the algebraic structure of the general solution, which is exactly the same
as for the well-known solutions with a translational Killing vector  which were originally constructed as lifts of 4D  solutions with vector multiplets \cite{Gaiotto:2005xt}:
\bea
 ds^2_5 &=& - f^2 ( dt + \xi )^2   + f^{-1} ds^2_4\label{fluxsol1}\\
\Theta^I &=& \left( -2 K^0 \star_3 d\left( { K^I\over K^0} \right) \right)^- \\
f^{-1} Y_I &=& - 2 K_I + D_{IJK} {K^J K^K \over K^0} \label{Yfeqs}\\
\xi &=& {\o \over 2} + { L\over 2 (K^0)^2} (d\theta^2 + \chi)\\
 L&=&K_0(K^0)^2+\frac{1}{3}D_{IJK}K^I K^J K^K -K^I K_I K^0\label{defJ}\\
F^I &=& - d( f Y^I ( dt + \xi )) + \Theta^I\label{fssol}
\eea
where we defined the (anti-) selfdual projections
\be
\a^\pm = \half (\a \pm \star_4 \a).
\ee
Recall that the orientation of the  base was chosen such that the K\"ahler form is selfdual.

In order to obtain  $f$ and $Y^I$ from (\ref{Yfeqs}), one has to solve the following quadratic equations for functions $y^I$
\be
 D_{IJK}y^I y^J=-2K_I K^0+D_{IJK}\,K^J K^K\
 \ee
from which $f$ and $Y^I$ are obtained as
\be
f = {2^{2/3} K^0\over \calq}, \qquad
Y^I=\frac{2^{1/3}y^I}{\sqrt{\calq}}, \qquad
 \calq=\left(\frac{1}{3}D_{IJK}y^I y^J y^K\right)^{2/3}.\label{Yfsol}
\ee

The examples we will consider in section \ref{secexs} below fall into a simple subclass of solutions where all the $K^I$ and all the $K_I$ are proportional to each other, $K^I = p^I\, \overline{K},\ K_I = p_I \underline{K}$, with
$p_I = D_{IJK} p^J p^K$. This ansatz leads in particular  to solutions where the
vector multiplet scalars $Y^I$ are constant:
\be
Y^I = \left({ 6 \over p^3} \right)^{1\over 3} p^I, \label{attrsols}
\ee
where we have defined
\be
p^3 \equiv D_{IJK} p^I p^J p^K\label{p3def}.
\ee
When the axidilaton is constant, this gives an attractor solution where the asymptotic moduli are fixed at their attractor values. Although not much is known about the attractor mechanism in the presence of hypermultiplets, it seems likely to us that such solutions are still attractors in the presence of the axidilaton.

Now let's discuss the differential equations which the various ingredients in the solution (\ref{fluxsol1}-\ref{fssol}) must satisfy.  The main difference with solutions with a translational Killing vector (such as those with a Gibbons-Hawking base)  is that the functions $K^0, K^I,K_I, K_0$ are not harmonic with respect to the 3D metric (\ref{3dmetrrot}) but instead satisfy
\bea
\D_{s_2} K^0&\equiv& d\star_3\left(dK^0+s_2(K^0)^2dy_2\right)=0\label{eqKu0}\\
\D_{s_2} K^I&\equiv& d\star_3\left(dK^I+s_2K^0K^Idy_2\right)=0\label{eqKuI}\\
\D_{s_2} K_I&\equiv& d\star_3\left(dK_I+s_2(K^0K_I+\frac{1}{2}D_{IJK}y^Jy^K)dy_2\right)=0\label{eqKlI}\\
\D_{s_2} K_0&\equiv& d\star_3\left(dK_0+\frac{s_2}{2}(K^IK_I-K^0K_0)dy_2+\frac{s_2}{2}\star_3(\omega\wedge dy_2)\right)=0\label{eqKl0}
\eea
while the one-forms $\chi,\o$ satisfy
\bea
\star_3 d \chi &=&  d K^0 + s_2 (K^0)^2 d y_2 \label{om0eq} \\
 \star_3 d\o & = & \langle d K,  K \rangle  -  s_2 L dy_2.\label{omeq}
 \eea
 where, in the second line, we have viewed  $K = (K^0, K^I,K_I, K_0)$ as a vector in a space equipped with a symplectic inner product
 \be
 \langle A, B \rangle = - A^0 B_0 + A^I B_I - A_I B^I  +   A_0 B^0
 \ee
In these expressions $\star_3 $ is  the 3D Hodge star taken with respect to the orientation  \linebreak $( \Re w^1, \Im w^1, y_2 )$.
The  equations (\ref{eqKu0}, \ref{om0eq})  follow from the Toda-like equation (\ref{Toda})  and the definitions (\ref{defKu0},\ref{defom0}). Note that we have introduced for later convenience the shorthand notation $\D_s$ to represent the differential operators acting on the $K$ functions; one should keep in mind that the action of $\D_s$ depends on which component of the symplectic vector $K$  it acts. We can then abbreviate (\ref{eqKu0}-\ref{eqKl0}) to
\be
\D_{s_2} K =0 \label{Keqs}.
\ee
Interestingly, the equations (\ref{eqKuI}-\ref{eqKl0}) are invariant under an $n_V$-parameter family of solution generating transformations:
\bea
K^I &\to& K^I + k^I K^0\\
K_I &\to& K_I + D_{IJK} k^J K^K + \half D_{IJK} k^J k^K K^0\\
K_0 &\to& K_0 + k^I K_I + {D_{IJK} \over 6} (k^I K^J K^K + 3 k^I k^J k^K K^0)\label{spectrflow}
\eea
with $k^I$ arbitrary real constants. We note that the quantity $L$ defined in (\ref{defJ}) is invariant under these transformations, so that from (\ref{omeq}) one easily  verifies that $\o$  is also invariant. This symmetry is a
generalization of the `spectral flow' symmetry in solutions with a translational Killing vector \cite{deBoer:2006vg},\cite{Cheng:2006yq},\cite{Bena:2005ni}.

Now let's discuss the integrability condition coming from (\ref{omeq}). Applying $d \star_3$ on both sides we find the condition
 \be
 0 = d d\o = \langle  \D_{s_2} K , K\rangle \label{integrab}
 \ee
 which is of course automatically satisfied if (\ref{Keqs}) holds with all components of  $K$  being smooth, i.e. without $\d$-function terms on the right hand side. More interesting is the  case is where one allows such $\d$-functions sources  and  replaces (\ref{Keqs}) with
 \be
 \D_{s_2} K = \sum_i \G_i \d^3(x -x_i) \mathrm{vol}_3 \label{deltasources}
 \ee
 Such singularities correspond in the  M-theory language to turning on, in addition to possible  M2-branes in the noncompact directions which source the axidilaton,  other  M-brane and momentum/KK monopole charges $\G$ at the positions $x_i$ on the 3D base which source the vector multiplets\footnote{ The $\calc_I$ represent a basis of two-cycles on the Calabi-Yau, with $\calc^I$ the dual basis of four-cycles with respect to the intersection product, and $S^1_{\theta^2}$ is the $\theta^2$ circle.}: $\G^0$  corresponds to a KK monopole, $\G^I$ to an M5 on $\calc^I\times S^1_{\theta^2}$, $\G_I$ to an M2 on $\calc_I$ and $\G_0$ to momentum along $S^1_{\theta^2}$. Note that when $s_2$ is zero these point charges are the only ones present, while in the case of $s_2\neq 0$ additional smeared charge may appear. It is very remarkable however, that even when $s_2\neq 0$ the well known stability equations \cite{Denef:2000nb, Bena:2005va} remain functionally the same. These equations  follow from the  the integrability condition (\ref{integrab}) which,  in the presence of    delta-function sources (\ref{deltasources}),   imposes nontrivial constraints on the allowed charges and positions of the branes in the 3D submanifold. Indeed, for each center we must impose
 \be
 \langle \G_i, K (x_i)  \rangle =0.\label{multicentconstr}
 \ee

When $\theta^2$ is periodic, the solutions with rotational Killing vector can be dimensionally  reduced to 4D along the $\theta^2$ direction, yielding an as yet  unexplored and  potentially  interesting class of multicentered solutions carrying various brane charges. One open question regarding such 4D solutions is whether  they are still supersymmetric. Since for a rotational Killing vector the 5D Killing spinor depends on $\theta^2$ (this follows from the $\theta^2$ dependence of $\F^\pm$ which are bilinears in the Killing spinor), it is not clear whether the
 reduced 4D solution will   preserve supersymmetry in general.
Furthermore, upon dimensional reduction one obtains a 4D metric which is a timelike fibration over
the 3D metric $ds^2_3$  given in (\ref{3dmetrrot}). Even for constant axidilaton, we know from (\ref{Ricci3d}) that this 3D metric is not flat in general. The reduced 4D metric  is then not obviously of the form introduced by Tod \cite{Tod:1983pm},\cite{Tod:1995jf}  which was shown to govern general supersymmetric solutions with  vector multiplets  \cite{Meessen:2006tu}. We feel that this interesting issue deserves further investigation.

The structure of the solutions simplifies  considerably when the Killing vector is translational, which can be obtained  as the limit $s_2\to 0 $ of the expressions above.
The 4D metric (\ref{4dmetrrot}) reduces to
\bea
ds^2_4 &=&  K^0 ds^2_3 + {1 \over K^0} \left( d \theta^2 +\chi   \right)^2\label{GHsol1}\\
ds^2_3 &=& {dy_2^2 } + \tilde \t_2 e^{\tilde h_1}  dw^1 d\bar w^1 \label{3dmetrtrans}
\eea
where, in view of (\ref{eqKu0}), $K^0$ is now a harmonic function of the 3D geometry, and  $\star_3 d\chi = d K^0$.
These metrics  are therefore generalizations of the Gibbons-Hawking metrics \cite{Gibbons:1979xm}, where the 3D base manifold is generically curved due to
the factor $\tilde \t_2$ in (\ref{3dmetrtrans}), see (\ref{Ricci3d}).
The form of the solution to the remaining equations determining the full 5D solution also simplifies significantly in the case of a translational Killing vector. The solution can still be written in the
form (\ref{fluxsol1}-\ref{fssol}), but now all functions are harmonic  in the  (generically curved) 3D metric (\ref{GH}), and the equation determining $\o$ also simplifies. Summarized, the equations (\ref{Keqs}),(\ref{omeq}) are now replaced by
\bea
 \nabla_3^2 K &=&0\\
 \star_3 d\o & = &    \langle K, d K \rangle.\label{GHsolfin}
 \eea
  When the axidilaton is constant, the 3D metric (\ref{3dmetrtrans}) becomes flat and the metric on the base is a Gibbons-Hawking hyperk\"ahler metric \cite{Gibbons:1979xm}. These solutions arise as
 5D uplifts  \cite{Gaiotto:2005gf}\cite{Gaiotto:2005xt},\cite{Behrndt:2005he},\cite{Cheng:2006yq},\cite{deBoer:2008fk}  of  the 4D N=2 vector multiplet solutions of \cite{Bates:2003vx}  describing type IIA multicentered  configurations of branes wrapped on the Calabi-Yau cycles. The
   constraints (\ref{multicentconstr}) are the well-known stability equations governing the existence of supersymmetric bound states \cite{Denef:2000nb}.
    A subset of these uplifted solutions are the 5D smooth bubbling geometries of \cite{Bena:2005va} containing topologically nontrivial cycles.

      Since in the case of a translational isometry this analysis has led to a wealth of insights in the BPS spectrum of string/M theory and phenomena such as wall-crossing \cite{Denef:2007vg}, it would  be of great interest to get a handle on  the constraints (\ref{multicentconstr}) on multicenter solutions with a rotational isometry. These may also play a role in constructing horizonless microstate geometries carrying the same charges as black holes or black rings (see e.g. \cite{Bena:2007kg} for a review and further references). In particular, the geometries in the black hole deconstruction proposal of \cite{Denef:2007yt} are multi-centered solutions of (\ref{eqKu0}-\ref{omeq}) with a rotational isometry and a nontrivial axidilaton from an M2 brane in the bulk. We will come back to this proposal in section \ref{secBHdeconstr}.

      Since the  hypermultiplets enter in (\ref{fluxeqs}) only implicitly through their impact on the 4D base metric, one would expect that the  results we derived in this section for  axidilaton solutions can be generalized  to  more general solutions involving hypermultiplets invariant under an  isometry.

We end this section by comparing our results to those obtained in \cite{Bellorin:2006yr}, which also considered supersymmetric solutions with an extra isometry under which $\t$ is invariant. In that work however, the Killing vector in question was tacitly assumed to be translational,  corresponding to the $s_2=0$ solutions (\ref{GHsol1}-\ref{GHsolfin}) in the present discussion, see in particular eqs. (4.83) in \cite{Bellorin:2006yr}. In uplifts of 4D  multicentered solutions, the Killing vector which generates translations on the M-theory circle is of the translational type. The restricted class of solutions considered in \cite{Bellorin:2006yr}  is suited to describe, for example,  backreacted M2 branes which are either localized or smeared on the M-theory circle. Our generalization (\ref{Toda},\ref{fluxsol1}-\ref{omeq}) on the other hand is needed  to describe configurations of M2 branes which are localized on the M-theory circle but which do possess a rotational Killing vector leaving $\t$ invariant. We will discuss an explicit example  in section \ref{Godelsec}. This class also contains the brane configurations in the black hole deconstruction proposal of \cite{Denef:2007yt}, on which we will comment in section \ref{secBHdeconstr}.

\subsubsection{Separated Toda solutions and enhanced symmetries}\label{septoda}
 In the case of a rotational Killing vector,  $s_2 \neq 0$, the  geometry of the  base is governed by the generalized Toda equation
(\ref{Toda}). Even when the axidilaton is constant, only few explicit solutions to this equation are known.
In this section we will concentrate on
 a special class of solutions 
to  (\ref{Toda}),
 where the Toda potential  is of the separated form
 \be
 e^{s_2 \Psi (w^1,\bar w^1, y_2)} =g(y_2) e^{- 2 \F (w^1,\bar w^1) }.\label{Todafact}
 \ee
 with $g(y_2)$ a positive real function.
For constant $\t$, this ansatz leads to the Liouville equation for $\F$  \cite{Gegenberg} and hence to  simple explicit solutions to (\ref{Toda}). We will find that also when $\t$ is turned on the separated ansatz
 leads to a more tractable subclass which includes  metrics with additional symmetries.
 We will now  derive these symmetry properties and give some explicit solutions to the equations (\ref{eqKu0}-\ref{omeq}) which lead  to highly symmetric 5D solutions.

With $e^{s_2 \Psi}$  of the factorized form (\ref{Todafact}), the equation (\ref{Toda}) implies a deformed Liouville equation for $\F$:
\be\begin{array}{ccl}
4 \F_{w^1\bar w^1}- \k^2 \tilde \t_2 e^{\tilde h_1} e^{-2 \F} &=& 0 \\
g '' &=& 2 \k^2\end{array} \label{Liouv}
\ee
where $\k^2$ is a real constant\footnote{By a shift of $\F$ we could set $\k^2$ to  either $-1,1$ or $0$, but we will find it convenient to keep $\k$ around.}.  Recall that $\tilde \t$ and $\tilde h$ are holomorphic functions of $w^1$.
The base metric is of the form  (\ref{4dmetrrot})  with
\bea
ds^2_4 &=&  K^0 ( dy_2^2 + g ds^2_2)  + {1 \over K^0} \left( d \theta^2 +
\chi \right)^2\label{4Dmetrsep} \\
K^0 &=& {g' \over s_2 g}; \qquad \chi =  {4 \over s_2} \Im (\F_{w^1} dw^1)\\
ds^2_2 &=&  \tilde \t_2 e^{\tilde h_1} e^{-2 \F} dw^1 d\bar w^1 \label{3Dmetrsep}
\eea

 Let's  first review the case when the axidilaton is constant \cite{Gegenberg}. Then the Liouville equation (\ref{Liouv}) implies that the 2D
 metric (\ref{3Dmetrsep}) has constant curvature, $R^{(2)}= 2 \k^2$.  Since 2D spaces of constant curvature are locally isomorphic to either the hyperbolic plane, the two-sphere or the plane depending on the sign of $\k^2$,  the solution has (locally) an additional three dimensional algebra of Killing vectors, namely $so(3)$, $sl(2,\RR)$ or the euclidean algebra $e(2)$ respectively. In fact, when $\k^2=0$, one can check that the base is completely flat and has local symmetry $e(4)$.

A natural question is thus whether it is possible to preserve this additional symmetry when the axidilaton becomes dynamic. Again,
for this to be the case it is necessary that the 2D metric (\ref{3Dmetrsep})  has constant curvature. Using the equation (\ref{Liouv}) one finds that the 2D curvature of a generic solution  is
\be
R^{(2)}= 2 \k^2 + { |\tilde \t'|^2 e^{2 \F} \over \tilde \t_2^3 e^{\tilde h_1}}\label{R2gensol} 
\ee
To have constant curvature  the second term should be a constant, which has to be determined  by demanding compatibility with (\ref{Liouv}). One finds that a constant curvature solution is possible only for negative $\k^2$ and is given by
\be
\F = 
\half \ln \left(- {  2 \k^2 \tilde \t_2^3 e^{\tilde h_1} \over 3 |\tilde \t'|^2 }\right), \qquad {\rm for\ }\k^2 <0
\label{GsolLiouv}.
\ee
for which the 2D metric (\ref{3Dmetrsep}) has curvature $R^{(2)}=  {4 \k^2 /3}$.
This solution to the deformed Liouville equation (\ref{Liouv}) was  considered in \cite{Levi:2009az} (and  in a different context in \cite{Kleban:2007kk}) and, as we will discuss in section \ref{Godelsec}, gives rise to 5D solutions which are (locally) G\"{o}del $\times$S$^2$. It is furthermore the unique solution to the  equation  (\ref{Liouv}) on a compact manifold without boundary \cite{trojanov}. However, as we  will see explicitly in section \ref{secBHdeconstr} and Appendix \ref{appdefliouv},
(\ref{GsolLiouv}) is  not the unique solution to (\ref{Liouv}) in the presence of a boundary, which will be the situation of interest to us. For the other solutions to (\ref{Liouv}) the
 2D metric $ds^2_2$ has non-constant  curvature (\ref{R2gensol})  and generically  doesn't possess  additional Killing vectors.

Let us now discuss how to construct full 5D solutions on a base determined by a separated Toda solution. First we should point out that the separated ansatz contains `ambipolar' base  metrics whose signature changes from mostly plus in one region to mostly minus in another. When this happens the 4D base is singular, as the metric eigenvalues pass through zero, but it is often possible to turn on  vector multiplet
fluxes  so as to give completely regular 5D metrics \cite{Bena:2005va}. From the form of the metric (\ref{4Dmetrsep}) we see that the metric is ambipolar  if and only if $K^0$ changes sign. It's easy to see that this can happen only when $\k^2 \neq 0$. In this case
by shifting $y_2$ with a constant we can assume  that the function $g$ is
\be
g=\k^2 y_2^2 + 4 a^2\ \qquad\mbox{when } \k^2\neq 0\label{gnonzerok}
\ee
with $a^2$ a real number. The range of $y_2$ must be chosen such that $g$ is positive. Hence we see  the base is ambipolar when $\k^2 \neq 0$ and $a^2>0$.
 Summarized, we have\bea
\k^2=0 \mbox{ or }  a^2<0<\k^2 &\Leftrightarrow& \mbox{ positive signature base}\,, \\
\k^2\neq 0,\, a^2 >0  &\Leftrightarrow& \mbox{ ambipolar base}\,.
\eea
Let us also note that  the  K\"{a}hler potential can be found explicitly by making a Legendre transform. For $g$ given by (\ref{gnonzerok}) the result is
\be
\calk = {4 a \over \k s_2} \left( \sqrt{{e^{s_2 x^2 + 2 \F }\over 4 a^2}-1}- \arctan \sqrt{{e^{s_2 x^2 + 2 \F }\over 4 a^2}-1} \right).\label{calkfact}
\ee
One checks that the Monge-Amp\`{e}re equation (\ref{MAKV}) is satisfied provided that (\ref{Liouv}) holds.

The next step in finding complete 5D solutions is to solve equations (\ref{eqKuI}-\ref{omeq}). When the base has positive signature, we can always trivially extend it to a five-dimensional static  solution with trivial vector multiplets by taking $K^I=K_I=K_0=\o=0$, so that
$ ds_5^2 = - dt^2 + ds^2_4$.

More interesting 5D solutions are obtained from ambipolar 4D bases. As remarked above, in this case we have to turn on vector multiplets if we want to have a chance of obtaining a regular 5D solution. It turns out that, from every  factorized solution (\ref{Todafact}) with ambipolar base it is possible to build a 5D solution (\ref{eqKuI}-\ref{omeq}) where either the timelike Killing vector $\pa_t$ (for  $\k^2 > 0$)  or the spacelike Killing vector $\pa_{\theta^2}$
(for $\k^2 < 0$) are part of an extended 3-dimensional symmetry algebra. The construction goes as follows.
Since $K^0= {g'\over s_2 g}$ depends only on $y_2$, it is natural to look for solutions where $K^I, K_I$ and $K_0$ also depend only on $y_2$.
Restricting attention to solutions of the form (\ref{attrsols}) which have constant vector multiplet scalars, and making suitable choices for the integration constants  for the resulting equations\footnote{See \cite{Bena:2007ju}, section 6.1 for a  discussion of the most general solution where $K$ depends only on $y_2$.}, one finds the following solution
to (\ref{eqKuI}-\ref{eqKl0}):
\be\begin{array}{lcllcl}
K^I &=&  { 2 a p^I \over  g}, &  K_I &=& - {   s_2^2 p_I\over 8 \k^2} K^0\\
K_0 &=& -{  s_2^2 a p^3 \over 12 \k^2} \left( {1 \over  g} - {b\over 2 a^2}\right), & L&=&  { p^3 \over 6 a g^2 }( 4a^2 + b \k^2 y_2^2)\\
\o &=& -{s_2^2 b p^3  \over 24 a \k^2} \chi + 2 \l, & d \l &=& 0 \label{specialKsol}
\end{array}\ee
The solution depends on the  free parameters $p^I $ (with $p^3$ as defined in (\ref{p3def})) and $b$ and on
a closed one-form $\l$. When $\l$ is exact, it can be absorbed in a redefinition of the time coordinate. A non-exact  one-form $\l$ will be
 needed to ensure that $\o$ has no Dirac string singularities and the integrability condition (\ref{integrab}) is satisfied, which  also has the effect of removing closed timelike curves \cite{Cheng:2006yq},\cite{Bena:2007ju}. We will discuss the required form of $\l$  in more detail in
section \ref{sectoricToda}.
The resulting 5D solution is
\bea
ds_5^2 &=& \left({p^3 \over 6}\right)^{2/3}\left[{dy_2^2 \over g} -{ g \over 4 a^2 \k^2} \a^2 +{1\over  \k^2} \left(\a  + {s_2 \over 2}( \chi+ d\theta^2) \right)^2 + \tilde \t_2 e^{\tilde h_1} e^{-2 \F} dw^1 d\bar w^1\right]\label{specialmetr}\\
  F^I &=& {p^I \over 2 a} d y_2 \wedge \a, \qquad  Y^I = \left({ 6 \over p^3} \right)^{1\over 3} p^I  \\
  \a &=& -{b s_2 \over 2} d\theta^2 - {24 a \k^2 \over s_2 p^3} (dt + \l),\qquad d \l =0  \label{special5dsol}
 \eea
with $g$ given in (\ref{gnonzerok}) with $\k^2 \neq 0,\, a^2>0$ and $\F$ a solution to (\ref{Liouv}).
Recall that on an ambipolar base $a$ is real, ensuring that the complete solution is indeed  manifestly real.
The first two terms in the metric (\ref{specialmetr}) constitute a 2D metric of curvature $-2 \k^2$, so that
for $\k^2 <0 $,  the solution contains an AdS$_2$ subspace fibered over a 3D base, while for
$\k^2 >0$ it contains a (fibered) round S$^2$. The vector multiplet flux is supported on  AdS$_2$ or S$^2$ respectively. The latter case is a `bubbled solution' in the spirit of \cite{Bena:2005va} and will play an important role later on. Hence we have demonstrated the existence of  bubbled solutions even in the presence of hypermultiplets.

\subsection{Solutions with toric K\"ahler base}\label{toricsection}
The discussion in the last subsection is as far as we managed to go for generic solutions with a single space-like Killing vector. Here we will see how demanding the presence of a second Killing vector, commuting with the first, constrains the solutions much more. The base becomes a (generalized) toric K\"ahler manifold,  and furthermore the possible profiles for $\tau$ are completely fixed by the symmetry up to some free constants and a few discrete choices. After working out those observations in the first subsection we illustrate them in the simple case with constant axidilaton, when the geometry is actually hyperk\"ahler. We then move on to dynamic axidilaton configurations, showing how the profiles we derived from symmetry have a physical interpretation as the presence of M2/exotic brane sources. Finally we use the separated solutions of the Toda equation to provide complete 5D solutions in the case with 2 Killing vectors.

\subsubsection{Toric K\"ahler manifolds and axidilaton profiles}\label{torK}
Having analyzed the structure of solutions with a single compatible (in the sense of section \ref{kvclass}) Killing vector, we can go one step further and impose that the 4D base has two commuting Killing vectors $k^{(1)}, k^{(2)}$. We choose  complex coordinates
\be
w^i = x^i + i \theta^i
\ee
which are adapted to the isometries, $k^{(i)} = {\pa \over \pa \theta^i}$, and
locally pick  a K\"ahler potential which is independent of both $\theta^i$:
\be
\calk = \calk (x^i).\label{toricK}
\ee
We then have, locally, from (\ref{killingpot})
\be
i_{k^{(i)}} \F_3 = - d y_i \label{mommaps}
\ee
with $y_i\equiv \calk_{x^i}$.
We will make here the extra technical assumption that the $y_i$ extend to globally well-defined functions, so that our Killing vectors are Hamiltonian and
 $y_i$ are the moment maps corresponding to $k^i$.  The  base
is then a toric\footnote{To be precise we should speak of a generalized toric manifold. The most conservative mathematical definition requires a 2$n$-dimensional compact symplectic manifold with a Hamiltonian $n$-torus action, so that the image of the moment map is a convex polygon by the  Atiyah-Guillemin-Sternberg theorem. This definition can be relaxed however to the non-compact setting, see e.g. \cite{Abreu}, where in return it is demanded that the moment map be proper onto its convex image, to ensure some polytopical form for that image. Here we will be a bit more loose in our nomenclature and for us a toric manifold will simply be any symplectic 2$n$ dimensional manifold with a Hamiltonian $\mathbb{T}^n$ action.} K\"ahler manifold. The restriction to toric 4D bases includes, as we will see below, a number of physically interesting
  situations and has the advantage of simplifying the BPS equations. In particular, the configurations we are most interested in will be governed by solutions of an ordinary nonlinear differential equation (see (\ref{Liouvtoric}) below). In addition, noncompact toric K\"{a}hler manifolds are a subject of recent interest in the mathematics community, see \cite{Abreu}
  and references therein.

Let's first discuss how the toric symmetry restricts the factor $\t_2 e^{h_1}$ in the Monge-Amp\`ere equation (\ref{defMA}). Repeating the analysis of section \ref{kvclass} for a second Killing vector, one finds that the function $\tilde \t$ in table \ref{table1} must be a linear function of $w^2$. Making a linear redefinition of the complex coordinates, we can arrange  that $\t$ depends only on $w^1$. In this coordinate frame $k^{(2)} = {\pa \over \pa \theta^2}$ leaves $\t$ invariant, while under the action of $k^{(1)} = {\pa \over \pa \theta^1}$, $\t$ is either invariant or transforms by a parabolic, hyperbolic or elliptic U-duality transformation. We will call these four cases  $II, IP, IH$ and $IE$ respectively.
The expressions for $\t$ and $h$ then reduce to (\ref{taugensol}) and  (\ref{h1KV})  with $\tilde \t (w^2) = \t_0 $   and
 $\tilde h (w^2) = s_2 w^2 + \ln c$, with $\t_0, c$ constants, so that they are fully determined by the symmetries. In what follows we will set $\t_0 =i$ and $c= 1,p,q,r$ in classes $II, IP, IH$ and $IE$ respectively. This choice will have the advantage that both $\t$ and $h$ remain well-defined in the limit $p,q,r \to 0$, a fact which will be useful later. We display the resulting expressions for $\t$ and $h$  with this choice of integration constants in table \ref{table4}.
\begin{table}\begin{center}\begin{tabular}{c|c|c|c}
Class & $\t(w^1)$&$h- s_i w^i$ & $e^{\m(x^1)} \equiv \t_2 e^{h_1- s_i x^i}$ \\ \hline
II &  $ i$ & $0$ &$  1$ \\ \hline
IP  &$ i(1 - p w^1 )$ &  $ 0 $&  $ 1- p x_1 $\\ \hline
IH & $i e^{- i r w^1 }$& $ i r w^1 $ &$\cos r x_1 $\\ \hline
IE &$ i \tanh (1- q  w^1 ) $& $ 2 \ln \cosh (1- q w^1)$&
$\half \sinh 2 (1- q x^1) $\\
\end{tabular}
\caption{The profiles for $\t, h$ and the function $\m$ for solutions with toric K\"ahler base.}\label{table4}
\end{center}
\end{table}
It will also be useful to single out the following combination, which depends only on $x^1$:
\be
e^{\m(x^1)}\equiv \t_2 e^{h_1 - s_i x^i} =\begin{cases} 1 & {\rm case \ II}\\
  {c\t_2 \over \sqrt{\t'\bar \t'}} &  {\rm cases \ IP,\ IH,\ IE}\end{cases}
\ee
Note that $\m $ solves a real Liouville equation:
\be
\m'' + c^2 e^{-2 \m}=0.\label{muliouv}
\ee
Explicit expressions for $\m$ are also given in table \ref{table4}.

The meaning of the constants $s_1,s_2$ is as follows. When both $s_1$ and $s_2$ are zero, both Killing vectors are translational. Note that when both $s_1$ and $s_2$ are nonzero, there is one linear combination of the Killing vectors
(namely $s_2 k^{(1)}- s_1 k^{(2)}$) which is translational while another combination is rotational. Therefore, from the moment that either $s_1$ or $s_2$ is nonzero we have one rotational
and one translational Killing vector.

We observe that in all cases  $  \t_2 e^{h_1}$ is independent of both $\theta^i$, which is consistent with the property (\ref{toricK}) and the equation (\ref{defMA}) for the K\"ahler potential.
In particular, in the toric case the equation (\ref{defMA}) becomes a real Monge-Amp\`ere equation:
\be
\calk_{x^1 x^1} \calk_{x^2 x^2}-(\calk_{x^1 x^2})^2 =   e^{\m + s_1 x^1 + s_2 x^2}\label{realMA}
\ee
and the 4D base metric can be written as
\be
 ds_4^2 = \calk_{ij} dx^i dx^j +   \calk_{ij} d\theta^i d \theta^j.
\ee

The toric K\"ahler geometry can also be described in terms of symplectic coordinates  $ y_i, \theta^i$ in terms of which the K\"ahler form takes the canonical (Darboux) form:
\be
\F_3 = dy_i \wedge d \theta^i
\ee
i.e. the $y_i$ play the role of canonical momenta conjugate to the torus coordinates $\theta^i$. The $y_i$ are the moment maps of the Killing vectors $k^i$ which, as we argued in (\ref{mommaps}),
are simply the derivatives of the K\"ahler potential:
\be
y_i = {\calk_{x^i}} \label{symplcoords}.
\ee
In symplectic coordinates the geometry  is encoded in  a  symplectic potential  $\cals$ which is related to the K\"ahler potential  by a Legendre transform with respect to $x^1$ and $x^2$:
\bea
\cals (y_i) &=& x^i y_i - \calk \label{symplpot}\\
 ds^2_4 &=& \cals^{ij} dy_i dy_j +  \cals^{-1}_{ij} d\theta^i d \theta^j, \qquad \cals^{ij} \equiv {\pa^2 \cals \over \pa y^i \pa y^j}
 \eea
 The coordinates $y_i$ trace out a convex region in $\RR^2$ called the {\em moment polytope}. It is determined by the requirement that
  \be
  \det \cals^{-1}_{ij} \geq 0.
  \ee
The edges of the moment polytope form the locus where the torus degenerates.
The symplectic potential also satisfies a Monge-Amp\`ere-type equation, namely
\be
\cals_{y_1y_1}\cals_{y_2y_2} - (\cals_{y_1y_2})^2 = { e^{- \m(\cals_{y_1},\cals_{y_2} ) -s_1 \cals_{y_1} - s_2 \cals_{y_2}} }\label{MAS}
\ee

There is a third description of the toric K\"ahler base which is the most useful for finding explicit solutions.  Since the axidilaton is left invariant by one of the Killing vectors, which in our conventions is $k^{(2)}= \pa_{\theta^2}$, we can describe the geometry in terms of a Toda potential  as
 we discussed in section \ref{invKV}. That discussion goes through unchanged in the toric case, the only difference being that all quantities are now independent of $\theta^1$.
 In particular, the Toda potential $\Psi (x^1,y_2)$ now satisfies a  Toda-like  differential equation in two real variables
\be
\Psi_{x^1 x^1} + { e^{\m+ s_1 x^1} \over s_2}  (e^{s_2 \Psi})_{y_2y_2} =0 \label{Todatoric}
\ee
The base  metric (\ref{4dmetrrot}) simplifies to
\bea
ds^2_4 &=&  K^0 ds^2_3 + {1 \over K^0} \left( d \theta^2 + \chi)  \right)^2\nonu
K^0 &=& \Psi_{y^2}, \qquad \chi = - \Psi_{x^1} d \theta^1\nonu
ds^2_3 &=& {dy_2^2 } +  e^{\m + s_1 x^1+ s_2  \Psi} \left((dx^1)^2 + (d\theta^1)^2\right)\label{basemetrtoric}
\eea
Given a solution to (\ref{Todatoric}) leading to a toric 4D base, one can look for  full 5D solutions preserving the toric isometries of the  base by solving the equations (\ref{eqKu0}-\ref{eqKl0}) for functions
 $K^I,K_I, K_0$  which are  independent of $\theta^1$.

\subsubsection{Toric hyperk\"ahler from Gibbons-Hawking} \label{secGH}
To illustrate the equations and solutions with toric symmmetry we first consider a class of examples with constant axidilaton, where  the 4D hyperk\"ahler base is a Gibbons-Hawking manifold.
Such solutions have a translational isometry, and we will look here at the subclass which has  an extra rotational   symmetry, so that the base is toric hyperk\"ahler with both a translational and a rotational Killing vector. This is the case  for example for a multi-Taub-NUT solution where all the centers lie on an axis, and  more generally any Gibbons-Hawking metric constructed from an axially symmetric harmonic function is a toric hyperk\"ahler metric. Although this is a known result
  in the mathematics literature \cite{Abreu}, we will rederive it here in a way that is completely explicit.
In such solutions, the base can be written both in the Gibbons-Hawking form, where the base is fibered along the translational direction over a flat 3D base, and the Toda form (\ref{basemetrtoric}), by taking the fiber to be the rotational $S^1$.
We derive here how these are related, and construct an  (implicit) solution to the Toda equation (\ref{Todatoric}) for every axially symmetric harmonic function. We then extend this correspondence to full 5D solutions, yielding an explicit non-trivial solution for the quantities $K, \o$ introduced in section \ref{mbsection} in terms of axisymmetric harmonic functions.

Our starting point is the well-known Gibbons-Hawking hyperk\"ahler metric, with an additional axial symmetry in the 3D flat base. In coordinates where $\pa_{\theta^1}$ generates the translational symmetry and $\pa_{\theta^2}$ generates the axial one the metric takes the form
\be
ds^2_4 = H^0 ( dr^2 + r^2 (d\theta^2)^2 + dz^2) + {1 \over H^0}  ( d \theta^1 + \tilde  H^0 d\theta^2  )^2\label{GH}
\ee
where $H^0$ is an axially symmetric harmonic function (depending only on $r$ and $z$). The function $\tilde  H^0 $ is defined as follows.  For any  axially symmetric harmonic function $H$ one can define a conjugate function
$\tilde H$ (see e.g. \cite{Weinstein}) satisfying
 $dH = \star_3 d( \tilde  H  d\theta^2 )$. In other words,
\bea
r \pa_r H &=& - \pa_z \tilde  H\label{harm}\\
r \pa_z H &=& \pa_r \tilde  H.\label{conjharm}
\eea
Integrability of these equations imposes that $H, \tilde  H$ satisfy the second order equations
\bea
\left( \pa_r^2 + r^{-1} \pa_r + \pa_z^2\right) H &=& 0\\
\left( \pa_r^2 - r^{-1} \pa_r + \pa_z^2\right) \tilde  H &=& 0.
\eea
It is not hard to check that $k^{(1)}=\pa_{\theta^1}$ is indeed a translational or triholomorphic Killing vector, which is equivalent to the associated one-form having self-dual curvature:
\be
dk^{(1)}=\star_4 dk\qquad k_a^{(1)}=g_{a\theta^1}.
\ee
This means this solution falls into the classification of section \ref{kvclass} with
\be s_1 = 0\,.
\ee

The second Killing vector $k^{(2)}=\pa_{\theta^2}$ is still holomorphic with respect to one of the complex structures, but no longer triholomorphic, making it rotational: $s_2\neq 0$. The analysis of section \ref{Tframe} then implies there should also exist a Toda form (\ref{basemetrtoric}) for the metric, which in this case is
\be
 ds^2_4 = \Psi_{y_2} \left( (dy_2)^2 + e^{s_2 \Psi} ( (d x^1)^2 + (d\theta^1)^2 \right) + {1\over \Psi_{y_2}}(  d \theta^2 -\Psi_{x^1} d\theta^1  )^2\label{Todaform}
\ee
As the two metrics (\ref{Todaform}) and (\ref{GH}) describe the exact same geometry we should be able to related them by a coordinate transformation. This might sound trivial, but it implies a rather intricate relation between solutions of the non-linear Toda equation and the simple linear Laplace equation in flat space. We start by equating the $d \theta^i d\theta^j$ terms in (\ref{GH}) and
(\ref{Todaform}), this gives the algebraic relations
\bea
e^{s_2 \Psi} &=& r^2\label{psiitor}\\
\Psi_{x^1}&=& -{ \tilde  H^0 \over (H^0)^2 r^2 + (\tilde  H^0)^2}\label{algeqs1}\\
\Psi_{y_2}&=&  { H^0 \over (H^0)^2 r^2 + (\tilde  H^0)^2}.\label{algeqs}
\eea
Note that $e^{s_2 \Psi/2}$ has the interpretation of the distance to the axis of symmetry in the flat metric.
Hence, to find the Toda potential, we simply have to solve for the radial distance $r$ in terms of the variables $x^1, y_2$.
Equating the remaining terms in (\ref{GH}) and
(\ref{Todaform}) then leads to the following relations between the coordinates:
\bea
dx^1 &=&-{\tilde  H^0 \over r} dr + H^0 dz\nonu
dy_2 &=& H^0 r dr + \tilde  H^0 dz.\label{x1y2}
\eea
Compatibility of these relations and (\ref{psiitor}) with  with equations (\ref{algeqs1},\ref{algeqs})  fixes
the rotational charge $s_2$ to be
\be
s_2 = 2.
\ee
A nice consistency check is to observe that (\ref{conjharm}) are exactly the conditions that the relations (\ref{x1y2}) can be integrated, i.e. they imply that $ddx^1 = ddy_2 =0$ so that $x^1, y_2$ are well-defined functions. Another interesting observation is that the  functions $x^1,y_2$  form a new pair of conjugate harmonic functions, since
\bea
r \pa_r x^1 &=& - \pa_z y_2\\
r \pa_z x^1 &=&  \pa_r  y_2
\eea
The couple $x^1, y_2$  is the `primitive' pair of  conjugate harmonic functions constructed from the pair $H^0,\tilde H^0$ \cite{Weinstein}. These relations let us also express the conditions (\ref{conjharm}) for a pair of functions  $(H, \tilde H)$ to be a harmonic pair, in the new coordinates:
\begin{eqnarray}\label{harcond2}
\partial_{x^1}H&=&\partial_{y_2}\tilde H\\
\partial_{x^1}\tilde H&=&-e^{s_2\Psi}\partial_{y_2} H
\end{eqnarray}
Using the relations (\ref{algeqs}) one can check that the above equations are equivalent to $\Psi$ satisfying the Toda equation (\ref{Toda}). 
Furthermore we have the relations
\bea
y_1 &=& z\nonu
x^2 &=& \ln r.\label{y1x2}
\eea

Let us also comment on how to find the K\"ahler and symplectic potentials. As discussed in section \ref{Tframe}, these can be obtained from either $H^0$ or $\Psi$ by integrating and making a Legendre transformation. In practice however, it is simpler to first find the relation between the K\"ahler coordinates $x^1, x^2$ and the symplectic coordinates $y_1,y_2$ through the
relations (\ref{x1y2}, \ref{y1x2}) and then integrate the equations
\bea
\calk_{x^1} &=& y_1 (x^1,x^2) \qquad \calk_{x^2} = y_2 (x^1,x^2)\\
\cals_{y_1} &=& x^1 (y_1,y_2) \qquad \cals_{y_2} = x^2 (y_1,y_2)\label{pots}
\eea

We can make these considerations more concrete for multi-centered Gibbons-Hawking bases, where all centers all lie on the $z$-axis:
\bea
H^0 &=& h^0 + \sum_i {q^0_i \over \sqrt{r^2 + (z-z_i)^2}}\label{H0GH}\\
\tilde H^0 &=& \sum_i {q^0_i (z-z_i) \over \sqrt{r^2 + (z-z_i)^2}}\\
x^1 &=&   h z + {1 \over 2} \sum_i q^0_i \ln { \sqrt{r^2 + (z-z_i)^2} + (z-z_i) \over \sqrt{r^2 + (z-z_i)^2} - (z-z_i)}\label{x1GH}\\
x^2 &=& \Psi = \ln r\\
y_1 &=& z\\
y_2 &=& { h r^2 \over 2} +  \sum_i q^0_i \sqrt{r^2 + (z-z_i)^2} \label{multicent}
\eea
In principle expressions for the Toda, K\"ahler and symplectic potentials can be found by inverting the relations above. But it looks hard, if not impossible, to explicitly relate the coordinates $x^i$ and $y_i$, in the case of more than two centers. Hence the implicit description above is as far as we can go in general.  For up to two centers with equal or opposite charges, an explicit description is possible and we will review these solutions in section \ref{secexs}.

Now that we have `solved' for the base in the Toda form we can go on and extend these relations to all quantities appearing in the full 5D supergravity solution discussed in section \ref{mbsection}. In `Gibbons-Hawking form' they are determined in terms of a number of harmonic functions\footnote{The $H$'s here are just the $K$'s of section \ref{mbsection} in the limit $s_2\rightarrow 0$, as the 'Gibbons-Hawking' frame is the one where we single out the translational Killing vector.} $H=(H^0, H^I, H_I,H_0)$ as
\bea
 ds^2_5 &=& - f^2 ( dt + \xi )^2   + f^{-1} ds^2_4 \label{GHbase1}\\
\Theta^I &=& \left( -2 H^0 \star_3^{GH} d\left( { H^I\over H^0} \right) \right)^- \\
f^{-1} Y_I &=& - 2 H_I + D_{IJK} {H^J H^K \over H^0}, \qquad D_{IJK} Y^I Y^J Y^K = 6\\
\xi &=& {\o^{GH} \over 2} + { L^{GH}\over 2 (H^0)^2} (d\theta^1 +\tilde H^0 d\theta^1 )\\
 \star_3^{GH} d\o^{GH} & = & \langle d H,  H \rangle \label{omGH}  \\
 L^{GH} &=&H_0(H^0)^2+\frac{1}{3}D_{IJK}H^I H^J H^K -H^I H_I H^0\\
F^I &=& - d( f Y^I ( dt + \xi )) + \Theta^I \label{GHbase}
\eea
When all harmonic functions are of the Coulomb form (\ref{H0GH}) with delta-function sources on the  $z$-axis, the integrability condition (\ref{multicentconstr}) for (\ref{omGH}) leads to the following constraint on the position of the centers:
\be
\sum_{j\neq i} {\langle q_i, q_j \rangle \over |z_i - z_j|} = \langle h, q_i \rangle\label{GHdistconstr}
\ee
for each of the centers labeled by $i$.

One can now find the expressions in the `Toda form' by comparing to the solution (\ref{fluxsol1}-\ref{fssol}) for non-zero $s_2$. Equating for example the two expressions for $\Theta^I$ one finds the relations
\bea
\pa_r\left( {K^I \over K^0} \right) &=& \tilde  H^0 \pa_r \left( {H^I \over H^0} \right) - r H^0 \pa_z \left( {H^I \over H^0} \right)\\
\pa_z\left( {K^I \over K^0} \right) &=& r H^0 \pa_r \left( {H^I \over H^0} \right) +\tilde  H^0 \pa_z \left( {H^I \over H^0} \right)
\eea
which are solved by
\be
K^I = {\tilde H^0 H^I - H^0 \tilde H^I \over (H^0)^2 r^2 + (\tilde  H^0)^2}\label{KitoH1}
\ee
where $\tilde H^I $ are conjugate harmonic functions for $H^I$.
Comparing the quantities $f^{-1} Y_I$ and $\x$ in both frames then determines the functions $K_I, K_0$ and the one-form $\o$ in the Toda form
of the solution:
\bea
K_I &=&  H_I - \half D_{IJK} \left({H^J H^K \over H^0 } -  {K^J K^K \over K^0 }\right) \\
K_0 &=&\Lambda^{GH}+\tilde H^0 (H^0)^{-2}L^{GH}+(K^0)^{-1}K^IK_I-\frac{(K^0)^{-2}}{3}D_{IJK}K^IK^JK^K \\
\o &=& \left((H^0)^{-2}L^{GH}+{ (K^0)^{-2}\left( (H^0)^2 r^2 + (\tilde  H^0)^2\right)^{-1}}  \tilde  H^0 L \right)d\theta_1\label{KitoH2}
\eea
where $\Lambda^{GH}$ is the solution to $d\Lambda^{GH}=\langle d\tilde H,H\rangle$.
It is an interesting exercise to check consistency by using the above expressions for $K$ in terms of $H$ to see that the equations (\ref{eqKu0}-\ref{eqKl0}) on $K$ reduce to the simple harmonic conditions (\ref{harcond2}) for $H$.

These solutions, which  arose from lifting well-known 4D multicentered solutions  to 5D, give rise to a
new and as yet unexplored class of 4D solutions  when we dimensionally reduce them along the rotational direction. This procedure   is a dimensionally reduced version of the `9-11 flip'
for type IIA/M theory solutions. It would be interesting to further study the 4D solutions obtained in this manner.

\subsubsection{Interpretation of toric solutions as  M2 (or exotic) branes }\label{intrM2}
We now proceed to discuss solutions on a toric base with non-constant axidilaton.
As we saw in section \ref{torK}, when we have toric symmetry  the axidilaton is completely fixed by symmetry, see  table \ref{table4}. One can now try to understand the physical interpretation of these profiles for $\tau$. This is most easily done in the case where the Killing vector under which $\t$ transforms non-trivially  generates a compact direction. In that case the axidilaton $\t$ has a nontrivial monodromy. Let us first consider the case that the monodromy is of parabolic type, which translates to a charge for the dual 4-form and hence the presence of  an M2-brane extended in the 5 external dimensions.  This can be generalized to the other classes of monodromy, but their interpretation in terms of basic M-theory objects is less understood, and we will refer to them as exotic branes following \cite{deBoer:2010ud},\cite{deBoer:2012ma}. Depending on the monodromy we will speak about
hyperbolic or elliptic branes. The parameters $p,q,r$ in table \ref{table4} then correspond to brane charges  and are quantized in string/M theory.  We summarize the relation between toric symmetries, $\t$ monodromies and brane sources in table \ref{table5}.
\begin{table}\begin{center}\begin{tabular}{c|c|c|c}
toric type & $\theta^1 \to \theta^1 + 2 \p$  monodromy  & brane type \\ \hline
IP & $\t \to \t + 2 \p p $& M2 (parabolic)\\
IH & $\t \to e^{2 \p r}\t $ & hyperbolic\\
IE & $\t \to { \cos 2 \p q \t + \sin 2 \p q \over - \sin 2 \p q \t + \cos 2 \p q}$ & elliptic
\end{tabular}\caption{Relation between toric symmetries, $\t$ monodromies and brane sources. The corresponding profiles for $\t$, $h$ and $\m$ can be read off from table \ref{table4}.}\label{table5}\end{center}\end{table}

In our notation, the Killing vector $k^{(2)}=\pa_{\theta^2}$ leaves $\t$ invariant while  $k^{(1)}=\pa_{\theta^1}$ induces a
U-duality transformation on $\t$. Hence in this convention $\pa_{\theta^2}$ acts along the worldvolume directions of the brane while $\pa_{\theta^1}$ acts in the transverse directions. We would like to interpret at least some of these M2 (or exotic) branes with toric symmetries as a backreaction of these branes in a given  background with toric base, such as the ones we reviewed in section \ref{secGH}. For this interpretation to work the solutions should be such that when taking the limit of zero charge $p,q,r\to 0$, the metric reduces to the desired background solution. The addition of the brane can preserve the toric symmetries of the background only if the brane is placed at a fixed point of $\pa_{\theta^1}$ in the 3D base; $\pa_{\theta^1}$ is then a rotational\footnote{Note that that this doesn't mean that $\pa_{\theta^1}$ must be of the rotational type in our classicifaction, i.e.
we can still have $s_1 =0$.}  symmetry when encircling the brane in the transverse space. Defining $u = |u| e^{i \theta^1}$ to be a local coordinate centered on the fixed point  of
$\pa_{\theta^1}$, the relation with the coordinate $w^1$ introduced earlier is
\be
u = e^{w^1}, \qquad x^1 = \log |u|.\label{ucoord}
\ee


One point we want to stress is that the solutions so obtained in general do not make sense globally,
since $\t_2$  can become negative in some part of the $u$-plane.
An analogous situation occurs in solutions involving D7 branes, where it is well-known that  in order  to make a globally well-defined solution one has to combine several such branes
\cite{Greene:1989ya},\cite{Bergshoeff:2006jj}, \cite{Braun:2008ua}. This comes at the cost of breaking the rotational symmetry in the transverse plane  and hence the toric character of the geometry.
The toric solution then  describes only the local geometry near one of the branes, as we will illustrate in example \ref{secm2flat}. A global non-toric example will
 be discussed in in example \ref{secm2glob}.
 An interesting loophole in the above argument arises when the locus where $\t_2$ becomes zero coincides with a boundary of the spacetime. This is in fact what happens in the examples we will discuss in sections \ref{Godelsec} and \ref{secBHdeconstr}.

\subsubsection{Separated toric solutions}\label{sectoricToda}
As in the case with one Killing vector, when $s_2 \neq 0$  a more tractable subset of toric  solutions is obtained by making a separated  ansatz for  the Toda potential.
 The analysis proceeds as in section \ref{septoda} with all quantities now independent of $\theta^1 = {\rm Im} w^1$.
 The geometry of the base is completely specified by a function $\F (x^1)$ which satisfies an ordinary nonlinear differential equation:
\be
\F''- \k^2 e^{\m + s_1 x^1-2 \F } = 0 \label{Liouvtoric}
\ee
where  $\m(x^1) $ can be read off from table \ref{table4}. The fact that the problem is reduced  to finding a function $\F$ satisfying an ordinary nonlinear differential equation is what makes this class of solutions (more) tractable and,  as we will see below in the examples, it
still contains a number of physically interesting solutions. The base metric (\ref{4Dmetrsep}) becomes
\bea
ds^2_4 &=&  K^0 ( dy_2^2 + g ds^2_2)  + {1 \over K^0} \left( d \theta^2 +
\chi \right)^2\nonu
K^0 &=& {g' \over s_2 g} \qquad \chi = {2 \over s_2} \F' \nonu
ds^2_3 &=& dy_2^2 + g  e^{\m + s_1 x^1-2 \F} \left( (dx^1)^2 + (d\theta^1)^2 \right) \label{3DmetrsepToda}
\eea

In our analysis in section \ref{septoda} we have already encountered solutions where the 4D base has extra symmetries. In particular, these possess two commuting Killing vectors and are toric. Let us write these solutions more explicitly in the current toric coordinates adapted to both isometries. For constant axidilaton (i.e. class $II$), $\m=0$ and we have the following solutions to
(\ref{Liouvtoric}) (up to some conveniently chosen integration constants):
\bea
\k^2 <0: & \F = \begin{cases}
 {s_1 x^1 \over 2} + \ln {2 \sinh {|\k| x^1 \over 2}} \\
 {s_1 x^1 \over 2} + \ln |\k|  x^1 \\
  {s_1 x^1 \over 2} + \ln {2 \sin {|\k| x^1\over 2} }\end{cases}\label{liouvk<0}\\
  \k^2 =0: & \F=0 \label{liouvk0} \\
\k^2 >0: & \F= {s_1 x^1 \over 2} + \ln { 2\cosh {\k x^1 \over 2} }\label{liouvk>0}
\eea
The different types of solution for $\k^2<0$ arise because in that case the base has an $sl(2,\RR)$ symmetry and we can choose $k^{(1)}$ to generate
an elliptic, parabolic or  hyperbolic isometry respectively.
When the axidilaton is turned on (i.e. in classes $IP,IH,IE$), we saw in (\ref{GsolLiouv}) that there exists special symmetric solution  for $\k^2<0$:
\be
\F = \half \ln\left(-  { 2 \k^2 e^{3 \m + s_1 x^1} \over 3 c^2}\right) \label{Gsoltoric}
\ee
where as before $c=p,q,r$ in classes $IP,\ IH$ and $IE$ respectively. To check that this in fact solves (\ref{Liouvtoric}) one has to use the property that $\m$ solves a Liouville equation (\ref{muliouv}).
The symmetries of the base manifold for these solutions were discussed in section \ref{septoda} and we recapitulate them in table \ref{table6}.
\begin{table}\begin{center}\resizebox{16cm}{!}{\begin{tabular}{c|c|c|c|c|c}
 class & $\k^2$ & $\F$ solution  & 4D  symmetry & 5D symmetry  & type \\ \hline
\multirow{6}{*}{II} & $<0$ &  (\ref{liouvk<0}) & $sl(2,\RR) \times u(1)$ & $sl(2,\RR)\times sl(2,\RR)\times so(3) $ &  AdS$_3 \times$S$^2$ \\\cline{2-6}
 & 0 & (\ref{liouvk0}) & $e(4) $& $iso(1,4)$ & $\RR^{1,4}$ \\  \cline{2-6}
 & $>0$&(\ref{liouvk>0})  & $so(3) \times u(1)$ & $a^2>0:\ so(4) \times sl(2, \RR)$  & AdS$_2 \times$S$^3$\\
&&&& $a^2<0:\ so(3) \times u(1) \times u(1)$  & Eguchi-Hanson$\times \RR$ \\ \hline
 \hline
\multirow{6}{*}{IP} & $<0$ & (\ref{Gsoltoric}) & $sl(2,\RR) \times u(1)$ & $sl(2,\RR)\times u(1) \times so(3) $ & Godel$\times$S$^2$\\
 &  & generic  (\ref{Liouvtoric}) & $u(1) \times u(1)$ & $u(1)\times u(1) \times so(3) $ & M2 in AdS$_3 \times$S$^2$ \\ \cline{2-6}
&$0$ & generic (\ref{Liouvtoric}) & $e(2) \times u(1)$ & $iso(1,2) \times u(1)    $  & M2 in $\RR^{1,4}$  \\ \cline{2-6}
&$>0$ & generic (\ref{Liouvtoric}) & $u(1) \times u(1)$ & $a^2>0:u(1) \times u(1)  \times so(3) $  & M2 in AdS$_2 \times$S$^3$ \\
&&&& $a^2<0: u(1) \times u(1)  \times u(1) $  & M2 in Eguchi-Hanson$\times \RR$ \\
\end{tabular}}
\caption{Factorized toric solutions  and their symmetries. The classes $IH,IE$ are similar to $IP$ but with the  M2-brane
replaced by an exotic brane.}\label{table6}\end{center}
\end{table}

We now discuss the full  5D solutions that we can construct on a toric K\"ahler base of the factorized form (\ref{3DmetrsepToda}), following the discussion in section \ref{septoda}. We distinguish three cases depending on the sign of $\k^2$ and $a^2$. The properties of these 5D solutions are summarized in table \ref{table6}, anticipating the more detailed discussion in section \ref{secexs}.\\
$\mathbf{\k^2 =0.}$ \\In this case the base has positive   definite signature
   and we can construct a static solution  with trivial vector multiplets.
The Toda equation is solved by taking $g = s_2 y_2 + 1, \F =0$, which leads to
\bea
ds^2_5 &=& - dt^2+  e^{\m+ s_1 x^1} \left( (dx^1)^2 + (d\theta^1)^2\right) + {dy_2^2 \over 1 + s_2 y2} + ( 1 + s_2 y2) (d\theta^2)^2\nonu
F^I &=& Y^I =0 .\label{flattype}
\eea
As we will discuss in more detail in examples \ref{secflat}, \ref{secm2flat} this type of solution contains flat $\RR^{1,4}$ (when the axidilaton is constant) and the near-brane geometry of an M2 or exotic brane in $\RR^{1,4}$.\\
$\mathbf{\k^2 >0, a^2<0.}$  \\
This base also  positive   definite signature
and  the static 5D solution looks as follows:
\bea
ds^2_5 &=& - dt^2 + {g'\over s_2} e^{\m + s_1 x^1 - 2 \F} \left( (dx^1)^2 + (d\theta^1)^2\right) + {g' \over s_2 g}dy_2^2 + {s_2 g \over g'} \left( d\theta^2 + {2 \over s_2} \F' d\theta^1\right)^2\nonu
F^I &=& Y^I =0\nonu
g &=&  \k^2 y_2^2 + 4 a^2, \qquad \k^2>0, a^2<0 \label{EHtype}
\eea
with $\F$ a solution to (\ref{Liouvtoric}).
As we will see in  examples \ref{secEH} and \ref{secBHdeconstr},
the solution with constant axidilaton is the $ \RR \times $Eguchi-Hanson metric while turning on the axidilaton allows us to describe
an M2 or exotic brane in this background.\\
$\mathbf{\k^2 \neq 0, a^2>0.}$  \\
In this case  the base is ambipolar and we can construct regular 5D solutions of the form (\ref{specialKsol}-\ref{special5dsol}) for the toric case, leading to
\bea
ds_5^2 &=& \left({p^3 \over 6}\right)^{2/3}\left[{dy_2^2 \over g} -{ g \over 4 a^2 \k^2} \a^2 +{1\over  \k^2} \left(\a  + \F' d\theta^1 +{s_2 \over 2} d\theta^2 \right)^2 +  e^{\m + s_1 x^1-2 \F} \left( (dx^1)^2 + (d\theta^1)^2\right)\right]\nonu
\a &=& -{b s_2 \over 2} d\theta^2 - {24 a \k^2 \over s_2 p^3} \left(dt - {R \over 2 } d\theta^1 \right)\label{5Dmetrtorambip}\\
  F^I &=& {p^I \over 2 a} d y_2 \wedge \a, \qquad  Y^I = \left({ 6 \over p^3} \right)^{1\over 3} p^I  \nonu
  g&=& \k^2 y_2^2 + 4 a^2, \qquad \k^2 \neq 0, a^2>0\nonu
 0&=&  \F_{ x^1 x^1}- \k^2  e^{\m + s_1 x^1 -2 \F}.\label{5Dambisol}
 \eea
 Here we have chosen the closed one-form in  (\ref{special5dsol}) to be proportional to $d \theta^1$: $\l = - R d \theta^1/2$.
As we will illustrate in examples \ref{secads3s2},\ref{secads2s3},\ref{Godelsec} and \ref{secBHdeconstr}, these solutions include the AdS$_3\times$S$^2$ and AdS$_2\times$S$^3$ backgrounds as well as M2/exotic branes added to them.

Let us now also discuss how to determine the constraint on the parameters that we have to impose in order for $\o$ to satisfy the integrability condition (\ref{integrab}) or, in other words, for it to be free of Dirac string singularities. Recall from (\ref{specialKsol}) that for the current class of solutions $\o$ is given by
\be
\o = -\left( {s_2 b p^3  \over 12 a \k^2} \F' + R \right) d \theta^1
\ee
If the range the coordinate is such that $|x^1|$ can become large,
the Liouville-like equation (\ref{Liouvtoric}) is  compatible with an asymptotically linear behaviour of $\Phi$ for large $|x^1|$:
\be
\F \sim  m |x^1|\label{Fas}
\ee
for some  constant $m$ large enough that the second in term (\ref{Liouvtoric}) vanishes for large $|x^1|$.
Hence in the 3D metric (\ref{3DmetrsepToda}), $\pa_{\theta^1}$ has a fixed line for $|x^1| \to \infty$, where there is a coordinate singularity.
The condition on $\o$ is then that the coefficient of $d\theta^1$ should vanish for large $|x^1|$, so that we should impose
\be
a = - ({\rm sgn} x^1) { s_2 b p^3 m   \over 12 R \k^2}\label{distconstrToda}
\ee
 We observe from (\ref{5Dmetrtorambip}) that for $\k^2<0$ this also removes closed timelike curves (CTCs) which would otherwise appear near the fixed line $|x^1|\to \infty$. As we will illustrate in
 example \ref{Godelsec}, it is still possible for CTCs to crop up  elsewhere in the spacetime, arising from having a lot of rotating matter.
  This is what happens in the classic example  of the G\"odel universe \cite{Godel:1949ga}, of which we will encounter a supersymmetrized version.

\section{Examples}\label{secexs}
In this section we will discuss some concrete examples of solutions with  a toric K\"ahler base. We begin by reviewing solutions with constant axidilaton where the 4D base is toric hyperk\"ahler, illustrating how the simplest solutions constructed from axially symmetric Gibbons-Hawking bases
as in section \ref{secGH} can also be obtained from separated solutions of the Toda equation of section \ref{sectoricToda}. We then turn to solutions with axidilaton, focussing on those solutions which describe backreacted M2 and exotic  branes placed in a background with toric base. We will discuss in detail backreacted branes in flat space and the highly symmetric G\"odel$\times$S$^2$ solution which, as we will argue, arises from a distribution of branes in the AdS$_3\times$S$^2$ background. We will also comment on the solutions describing individual branes in the AdS$_3\times$S$^2$, Eguchi-Hanson and AdS$_2\times$S$^3$ backgrounds, which will be discussed in more detail in a separate publication.
\subsection{Solutions with toric hyperk\"ahler base}\label{hyperkex}
We have discussed two ways of constructing such solutions: from
axisymmetric solutions with Gibbons-Hawking base in section \ref{secGH} and from solving the Toda equation with a separated ansatz in section \ref{sectoricToda}.
We will see that  examples of the first method with up to two Gibbons-Hawking centers
fit precisely  in the second type of solutions with one translational  and one rotational Killing vector.
\subsubsection{Flat spacetime}\label{secflat}
To get a feeling for the definitions and coordinate systems introduced above, let's warm up by seeing how the flat $\RR^{1,4}$ background fits in our formalism. The simplest solution to (\ref{realMA}) with $\m=0$ is of the separated form
\be
\calk = {e^{s_1 x^1} - s_1x^1-1 \over s_1^2} + {e^{s_2 x^2} - s_2 x^2-1 \over s_2^2}
\ee
The linear terms, which could be removed by  a K\"ahler transformation, are added to have a well-behaved $s_i \to 0$ limit.
The corresponding 5D solution with trivial vector multiplets gives the Minkowski metric in the form
\be
ds^2_5 = -dt^2 + e^{s_1 x^1} \left( (dx^1)^2 + (d\theta^1)^2\right)+ e^{s_2 x^2} \left( (dx^2)^2 + (d\theta^2)^2\right).\label{flatmetr}
\ee
This example illustrates the origin of the terminology of translational and rotational Killing vectors: for $s_i\to 0$, the Killing vector $\pa_{\theta^i}$ generates a Minkowski translation
while for $s_i >0  $ it generates a rotation around the fixed point of $\pa_{\theta^i}$ which is at $x^i\to - \infty$. For the minimal quantum $s_i=2$ the space is free of conical singularities\footnote{Recall that, in case the $\theta^i$ is a compact coordinate, we have chosen its  period to be $2 \p$.}    while $s_i = 2 N$ gives a $\ZZ_N$ orbifold singularity in the fixed point.

From (\ref{symplcoords}) we find the symplectic coordinates $y_i$:
\be
x^1 = {\ln ( 1 + s_1 y_1 )\over s_1}, \qquad x^2 = {\ln ( 1 + s_2 y_2 )\over s_2}.
\ee
In particular, the Toda potential is
\be
\Psi = x^2 = {\ln ( 1 + s_2 y_2 )\over s_2}
\ee
which corresponds to a factorized solution (\ref{Todafact}) of the Toda equation (\ref{Todatoric}) with $g= s_2 y_2 +1$ and $\F=0$. In the Toda frame,
the metric is precisely (\ref{flattype}) with $\m=0$.

Making the Legendre transform (\ref{symplpot}) we find the symplectic potential
\be
\cals = { (1+ s_1 y_1)\log(1+ s_1 y_1)- s_1 y_1 \over s_1^2} + { (1+ s_2 y_2)\log(1+ s_2 y_2)- s_2 y_2 \over s_2^2}
\ee
which indeed satisfies the Monge-Amp\`ere-type equation (\ref{MAS}). From the form of the metric  in symplectic coordinates
\be
ds^2_5 = -dt^2 + {dy_1^2 \over 1 + s_1 y_1}+  {dy_1^2 \over 1 + s_1 y_1} + (1 + s_1 y_1) (d\theta^1)^2 + (1 + s_1 y_1) (d\theta^1)^2
\ee
one sees that the moment polytope of flat space is
\be
y_1 \geq -{1 \over s_1}, \qquad y_2 \geq -{1 \over s_2}.
\ee

For $s_1=0, s_2=2$ we can alternatively describe the solution in terms of an axially symmetric Gibbons Hawking base (\ref{GHbase1}-\ref{GHbase}) where
\be
H^0 = 1,\qquad r = \sqrt{1 + 2 y_2}= e^{x^2},\qquad z = x^1 = y^1
\ee
and all the other quantities $\tilde H^0 = H^I = H_I =H_0 = \o^{GH}$ are taken to be  zero.  As a check one easily verifies that the relations (\ref{algeqs}) are indeed satisfied.

\subsubsection{Two-center Taub-NUT: Eguchi-Hanson}\label{secEH}
\begin{figure}
\begin{center}
\begin{picture}(140,90)
\put(-100,0){\includegraphics[height=90pt]{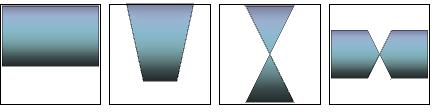}}
\end{picture}\end{center}
\caption{The moment polytopes for (a) flat space with $s_1=0, s_2 \neq 0$, (b) the Eguchi-Hanson metric,
(c) the ambipolar continued Eguchi-Hanson metric, and (d) the ambipolar metric for a Taub-NUT anti-Taub-NUT system.}\label{pols}
\end{figure}

Next, let's consider a Gibbons-Hawking base with two centers of equal charge at a distance $2b$ apart  and without constant term in $H^0$ (i.e.  ALE):
\be
H^0 = {P \over 2 \sqrt{r^2 + (z-b)^2}}+  {P \over 2 \sqrt{r^2 + (z+b)^2}}
\ee
As we will see, $P$ the is NUT charge of the solution (or the D6-charge after dimensional reduction along the translational direction), and for  $P=1$ the metric on the base is  the  Eguchi-Hanson metric \cite{Eguchi:1978xp}.
Since this is axially symmetric we can rewrite it in a toric form with one rotational and one translational Killing vector. To do this we first work out the relation between complex and symplectic coordinates using (\ref{x1GH}-\ref{multicent}):
\bea
x^1 &=& P{\rm arctanh} {P y_1\over y_2}\\
x^2 &=& \ln {\sqrt{(y_2^2-  b^2 P^2) (y_2^2 - P^2 y_1^2)}\over P y_2}
\eea
The  Toda potential can be computed from (\ref{pots}) and one  finds that it is of the factorized form (\ref{Todafact}) with  with positive $\k^2= {4/P^2}$:
\be
e^\F = 2 \cosh {x_1\over P}\qquad g = 4\left( {y_2^2\over P^2} - b^2\right)
\ee
From our discussion in section \ref{septoda} we know that the solution has  $so(3)\times u(1)$ isometry, which is indeed a well-known property of the Eguchi-Hanson metric.
The K\"ahler potential is given by (\ref{calkfact}),
 which can be brought into the more  standard form which makes the $so(3)\times u(1)$ symmetry manifest by making a holomorphic coordinate transformation
\be
u^\pm = {1\over \sqrt{2}} e^{\half( w^2 \pm {w^1\over P})}\label{upm}
\ee
The K\"ahler potential then becomes
\be
\calk = b P\left( \sqrt{ 1 + {(u^+ \bar u^+ + u^- \bar u^-)^2 \over b^2}} - {\rm arctanh} \sqrt{ 1 + {(u^+ \bar u^+ + u^- \bar u^-)^2  \over b^2}}\right)
\ee
We can also compute the symplectic potential from (\ref{pots}):
\be
\cals =  y_2\ln {\sqrt{ (y_2^2-  b^2 P^2) (y_2^2 -  P^2 y_1^2)}\over  y_2 P}-  b  P{\rm arctanh} { b P\over y_2}+ y_1 P {\rm arctanh} {P y_1\over y_2}  - y_2\nonu
\ee
The corresponding moment polytope is the following region in $\RR^2$ (see figure \ref{pols}(b)):
\be
 |y_2| \geq  P b, \qquad |y_2| \geq P |y_1|.
\ee

It's convenient to introduce prolate spheroidal coordinates:
\bea
r &=& b \sinh \r \sin \h\\
z  &=&  b \cosh \r \cos \h\label{prolspher}
\eea
in terms of which $x^1, y_2$ are
\bea
x^1 &=& P \ln \cot {\h \over 2}\\
y_2 &=&  b P \cosh \r.
\eea
The static 5D metric constructed from this 4D base is then  (\ref{EHtype})  with $ s_1=0, s_2=2, \k^2 = 4/P^2, a^2 = - b^2$ and takes the form
\be
ds^2_5 =- dt^2 +   b P \cosh \r \left( d\h^2 + {1\over P^2} \sin^2 \h (d\theta^1)^2 + d\r^2 + \tanh^2 \r( d\theta^2 + {1\over P} \cos \h d\theta^1)^2\right)
\ee
Note the manifest $so(3) \times u(1) \times u(1) $ symmetry.
For generic value of $P$, there will be a conical singularity at $\r =0$, but for $P=1$ we obtain the smooth Eguchi-Hanson manifold.
  In the limit $b\to 0$, the solution reduces to the geometry near a   single centered Taub-NUT  charge $P$, which for $P=1$ is equivalent to flat spacetime.

\subsubsection{The AdS2 x S3 solution}\label{secads2s3}
As a first example of a solution with an ambipolar base, consider the base of the previous example upon the analytic continuation   $b \to i a$.
 For $P=1$ the resulting metric  was already discussed in \cite{Eguchi:1978xp} where it was called  `type II'.
 The Toda potential is of the factorized form (\ref{Todafact}) with  with positive $\k^2= {4/P^2}$ and
\be
e^\F = 2 \cosh {x_1\over P}\qquad g = 4\left( {y_2^2\over P^2} + a^2\right).\label{Fads2s3}
\ee
The  K\"ahler potential is, in the variables (\ref{upm}):
\be
\calk
= a P \left( \sqrt{  {(u^+ \bar u^+ + u^- \bar u^-)^2 \over a^2}-1} - {\rm arctan} \sqrt{  {(u^+ \bar u^+ + u^- \bar u^-)^2 \over a^2}-1}\right)
\ee
which shows the $so(3)\times u(1)$ isometry discussed in section  \ref{septoda}.
The  symplectic potential is
\be
\cals =  y_2\ln {\sqrt{ (y_2^2+  a^2 P^2) (y_2^2 - P^2 y_1^2)}\over  y_2 P}-  a  P{\rm arctan} { a P\over y_2}+ y_1 P {\rm arctanh} {P y_1\over y_2}  - y_2.
\ee
The image of the moment map is a conical region in $\RR^2$ (see figure \ref{pols}(c)):
\be
|y_2|\leq P |y_1|
\ee
This  ambipolar metric is therefore characterized by a generalized moment  polytope which consists of two convex regions, the upper and lower parts of the cone,
 whose tips touch at $y_1=y_2=0$, where the base becomes singular.

After the change of coordinates
\bea
x^1 &=& P \ln \cot {\h \over 2}\\
y_2 &=&  b P \sinh \r.
\eea
the metric on the 4D base is manifestly ambipolar:
\be
ds^2_4 =  a P \sinh \r \left( d\h^2 + {1\over P^2} \sin^2 \h (d\theta^1)^2 + d\r^2 + \coth^2 \r( d\theta^2 + {1\over P} \cos \h d\theta^1)^2\right).
\ee
The fact that the base is ambipolar means that we will have to turn on vector multiplets  to get a regular 5D solution.
We take the solution of the form (\ref{5Dmetrtorambip}), where in the present example we should take $s_1=0, s_2=2, k^2 = 1/Q^2, a^2 >0$ and $\F$ given by (\ref{Fads2s3}).
Taking furthermore the parameters $b = R=0$ in   (\ref{5Dmetrtorambip}) gives
$\o=0$ and hence no further constraints  have to be imposed on the parameters from regularity of $\o$.
The resulting solution is
\bea
ds^2_5 &=& {P^2 \over 4} \left({p^3 \over 6}\right)^{2\over 3}\left[ - \left( {24 a \cosh \r \over p^3 P} \right)^2 dt^2 + d\r^2 \right. \nonu
&& \left.+ d\h^2 + d\left({\theta^1 \over P}\right)^2+ d\left( \theta^2 - {48 a  \over p^3 P^2} t \right)^2 + 2 \cos \h d\left({\theta^1 \over P}\right) d\left( \theta^2 - {48 a  \over p^3 P^2} t \right)\right]\\
F^I &=& - p^I \left( {24 a \cosh \r \over p^3 P} \right) dt\wedge d\r, \qquad Y^I = \left({ 6 \over p^3} \right)^{1\over 3} p^I
\eea
The metric is locally AdS$_2 \times$S$^3$ with a spinning S$^3$ and with flux on AdS$_2$. This solution  arises as the near-horizon limit of charged BPS black holes.
\subsubsection{TN-anti-TN with flux}\label{secads3s2}
An important solution with ambipolar base comes from a Gibbons-Hawking metric with two oppositely charged centers:
\be
H^0 = {P \over 2\sqrt{r^2 + (z+a)^2}}-  {P \over 2\sqrt{r^2 + (z-a)^2}}
\ee
Working out the relation between complex and symplectic coordinates  one finds
\bea
x^1 &=& P{\rm arccoth} {P  y_1\over y_2}\\
x^2 &=& \ln {\sqrt{(y_2^2-  a^2 P^2 ) (y_2^2 - P^2 y_1^2)}\over P y_2}
\eea
The Toda potential $\Psi$ is of the factorized form (\ref{Todafact}) with negative $\k^2= -4/P^2$:
\be
e^\F = 2 \sinh {x_1\over P},\qquad  g = 4\left( a^2 -{y_2^2 \over P^2}\right)\label{F66b}
\ee
so that we know from the discussion in section \ref{septoda} that the 4D base has
$sl(2)\times u(1)$ isometry. The Kahler potential is, in the variables (\ref{upm}):
\be
\calk =  a P\left( \sqrt{ 1 - {(u^+ \bar u^+ - u^- \bar u^-)^2 \over a^2}} - {\rm arctanh} \sqrt{ 1 - {(u^+ \bar u^+ - u^- \bar u^-)^2  \over a^2}}\right).
\ee
For the symplectic potential one finds
\be
\cals =   y_2\ln {\sqrt{ (y_2^2- a^2 P^2) (y_2^2 - P^2 y_1^2)}\over P y_2}-  a P{\rm arctanh} { a P\over y_2}-P y_1 {\rm arctanh} {P y_1\over y_2}  - y_2\nonu
\ee
The image of the moment map  is in this case the region
\be
|y_2| \leq a P, \qquad |y_2|\leq P |y_1|\label{pol66b}
\ee
which once again consists of two convex regions touching in the point $y_1=y_2 =0$ where the base is singular, see figure \ref{pols}(d).

It is once again convenient to switch to prolate spheroidal coordinates (\ref{prolspher})
in terms of which
\bea
x^1 &=& P \ln \coth {\r \over 2}\\
y_2 &=&  a P \cos \h
\eea
and the ambipolar  base metric is manifestly $sl(2,\RR)\times u(1)$  symmetric:
\be
ds^2_4 =  -{ a P} \cos \eta  \left(d \rho^2  +\frac{1}{P^2} \sinh ^2\rho
  (d \theta^1)^2 +d\eta^2 +\tan^2\eta  \left( d\theta^2 +\frac{1}{P} \cosh \rho
   d \theta^1\right)^2\right)
\ee

We can construct a regular 5D solution on this base  by turning on appropriate fluxes. Indeed, it arises from lifting  a D6-anti D6 system
which can  be made stable only if suitable worldvolume fluxes are turned on, providing additional repulsive forces.
In the Gibbons-Hawking form, the relevant solution is of the form (\ref{GHbase1}-\ref{GHbase}) with harmonic functions \cite{Denef:2007yt},\cite{deBoer:2008fk}:
\bea
H^I &=&{p^I P\over 4} \left( {1 \over \sqrt{r^2 + (z+a)^2}}+ {1 \over \sqrt{r^2 + (z-a)^2}}\right)=- {p^I P\over 2 a} {\cos \h \over \cosh^2 \r -\cos^2 \h }\\
H_I &=&{p_I P\over 16} \left( {1 \over \sqrt{r^2 + (z+a)^2}}-  {1 \over \sqrt{r^2 + (z-a)^2}}\right)={p^I P\over 8 a} {\cosh \r \over \cosh^2 \r -\cos^2 \h }\\
H_0 &=&{p^3 P\over 96} \left( {1 \over \sqrt{r^2 + (z+a)^2}}+  {1 \over \sqrt{r^2 + (z-a)^2}}\right)- {R }= {p^3 P\over 48 a} {\cos \h \over \cosh^2 \r -\cos^2 \h }-R \nonu
\eea
where the parameter $R$ is related to the asymptotic radius of the M-theory circle.

Rewriting the solution in the Toda form using the formulas (\ref{KitoH1}-\ref{KitoH2}), one finds that the functions $K$ for this solution are  precisely of the form (\ref{specialKsol}) where the parameter $b=1$.
It is also instructive to see how the constraint   on the distance $2 a$ between the centers arises in both frames. In the Gibbons-Hawking
frame, the constraint is (\ref{GHdistconstr}) which gives
\be
a = {p^3 P \over 24 R}.\label{distd6ad6}
\ee
From the point of view of the Toda frame, the same constraint follows from equation (\ref{distconstrToda}) imposing absence of  singularities
in $\o$ . From the $x^1 \to \infty$ behavior
of $\F$, we see that the constant $m$ in (\ref{Fas}) is $m = {1 \over P}$, leading once again to (\ref{distd6ad6}).

Using (\ref{distd6ad6})  to eliminate $a$ , we obtain the  full 5D solution
\bea
ds^2_5 &=& {P^2\over 4} \left({p^3 \over 6}\right)^{2\over 3}\left[ - \left( {2 dt \over P R} + ( \cosh \r -1) {d\theta^1 \over P} \right)^2 + d\r^2 + {1\over P^2} \sinh^2 \r (d \theta^1)^2 \right. \label{ads3s2metr}\\
&& + d\h^2 + \sin^2\h d\left(\theta^2 + {\theta^1 \over P}- {2 t \over P R}   \right)^2\Big]\\
F^1 &=& - {p^I P\over 2} \sin \h d\h  \wedge d\left(\theta^2 + {\theta^1 \over P}- {2 t \over P R}   \right), \qquad Y^I = \left({ 6 \over p^3} \right)^{1\over 3} p^I
\eea
We again obtain a smooth solution for $P=1$, corresponding to precisely one unit of NUT charge at each of the centers.
The first part of the metric is then global AdS$_3$ in rotating coordinates, written as a fibration over the hyperbolic plane. The second part is a (nontrivially fibered) round S$^2$ or a `bubble' in the language of \cite{Bena:2005va}.


\subsection{Solutions with toric K\"ahler base}
In the remainder of this section we will consider examples where the axidilaton is turned on and which have a toric K\"ahler base.
Following the general discussion in section \ref{intrM2} we will focus on solutions which describe the backreaction of   M2- or exotic branes placed in one of the backgrounds with hyperk\"ahler
base that were discussed in the previous section. All these solutions fit within the separated toric ansatz discussed in section \ref{sectoricToda}.

\subsubsection{Local M2-branes in flat space}\label{secm2flat}
We start by studying the solutions describing the local geometry near M2- or exotic branes in flat space following the procedure
described in section \ref{intrM2}.
Consider flat space parameterized as in (\ref{flatmetr}) with $s_1=2$ and $s_2=0$, so that the Killing vector $\pa_{\theta^1}$  is rotational with $s_1=2$ while
$\pa_{\theta^2}$ is translational. The  Killing vector $\pa_{\theta^1}$ has a fixed locus of codimension 2 at $x^1 = - \infty$ and, as
discussed in section \ref{intrM2}, we can backreact an M2- or exotic brane there without spoiling the toric symmetry. The resulting solution is given by  (\ref{flattype}) with   $s_1=2$ and $s_2=0$. The Toda potential is simply  $\Psi = y_2$ and transforming to complex coordinates one finds the K\"ahler potential $\calk =   M(x^1)+ {(x^2)^2 \over 2}$ where the function $M$ should satisfy
\be
 M'' = e^{\m + 2 x^1}.
\ee
The full solution is
\bea
ds_5^2 &=& - dt^2 + e^{\m(x^1) + 2 x^1} dw^1 d \bar w^1   + dw^2 d \bar w^2\\
&=& - dt^2 + e^{\m(\ln|u|) } du d \bar u   + dw^2 d \bar w^2
\eea
where in the second line we have switched to  the local coordinate (\ref{ucoord}) $u = e^{w^1}$ centered on the brane position.
The curvature $R^{(2)}$ of the transverse space and the accumulated  deficit angle $\d$ as a function of $|u|$ are given in terms of $\m$ by
\bea
R^{(2)} &=& {c^2 e^{- 3 \m (\ln |u| )} \over |u|^2}, \qquad c = p,q,r\\
\d (|u|) &=& - \p   \m' (\ln |u| )\label{curvdelta}
\eea
We will now discuss the resulting local geometry near each type of brane in turn.

\begin{figure}
\begin{center}
\begin{picture}(200,70)
\put(-50,0){\includegraphics[height=90pt]{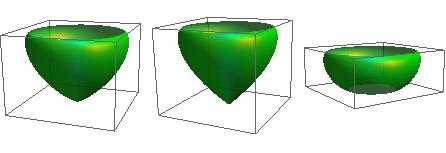}}
\end{picture}\end{center}
\caption{The geometry transverse to (a) an M2-brane, (b) an elliptic brane and (c) a hyperbolic brane, illustrated as a (local) isometric embedding in $\RR^3$.}\label{M2fig}
\end{figure}

For the M2 brane, which has parabolic  monodromy $\t \to \t + 2\p q$ under $u \to e^{2 \p i} u$, we obtain
\bea
\t(u) &=& i \left( 1 - {p \ln u }\right); \qquad h(u)=0\\
ds_5^2 &=&-dt^2 +  \left(1 -  {p \ln |u| }\right) {du d\bar u}  + dw^2 d \bar w^2.
\label{flatM2}
\eea
We see from (\ref{curvdelta}) that there is a mild, integrable, curvature singularity at the brane position $u=0$, although there is no conical deficit there as was observed in \cite{Bergshoeff:2006jj},\cite{Braun:2008ua}.
There is a further singularity at $|u| = e^{1/p}$ where also $\t_2$ vanishes and the solution breaks down. This illustrates  the fact that the solution with only a single M2-brane does not make sense
globally and other objects need to be introduced to make the solution well-behaved. We will illustrate how to do this in the next section.
The geometry of the  space transverse to the M2 brane, isometrically embedded in $\RR^3$,  is illustrated in figure \ref{M2fig}(a).

For the elliptic brane, which has monodromy $\t \to { \cos 2\p q \t + \sin 2\p q \over - \sin 2\p q \t + \cos 2\p q}$ under $u \to e^{2 \p i} u$, the local geometry is
\bea
\t(u) &=& i \tanh ( 1 - q \ln u) ; \qquad h(u)=2 \ln \cosh (1 - q \ln u)\\
ds_5^2 &=& - dt^2 +\half \sinh 2( 1- q \ln |u| )  du d\bar u   + dw^2 d \bar w^2\label{ellsol}
\eea
One finds that such a brane produces a conical singularity with deficit angle $\d (0) = 2 \p q$ at the brane position, see  figure \ref{M2fig}(b).
Once again there is a curvature singularity at $|u| = e^{1/p}$ where the local solution breaks down.

The hyperbolic brane has monodromy $\t \to e^{2\p r} \t $ and leads to the local geometry 
\bea
\t &=& i u^{-{i r}}; \qquad h= { i r  \ln u}\\
ds_5^2 &=&- dt^2 + \cos  {r \ln |u| } { du d\bar u }  + dw^2 d \bar w^2
\eea
Since the solution is periodic in $\ln |u|$ it is problematic to interpret it as arising from a single brane source at $u=0$\footnote{In
 type IIB/F-theory, the status of the object with hyperbolic monodromy is similarly unclear \cite{Bergshoeff:2006jj}.}. It is natural to restrict $|u|$ to a single period
\be
e^{- {\p \over 2 r}} \leq |u| \leq e^{ {\p \over 2 r}};
\ee
since there are curvature singularities  at both ends of the interval. See  figure \ref{M2fig}(c).

 As an extra check we see that we indeed recover the flat metric when we let the charges $p,q,r$ go to zero (or equivalently, let $u$ approach $|u|=1$).
We recognize in these formulas the 5D versions of the local  backreacted   `Q-brane' solutions in flat space   of \cite{Bergshoeff:2006jj}.

\subsubsection{Global M2-brane  solutions}\label{secm2glob}
We saw that solutions containing a single M2 or exotic brane in flat space do not make sense globally as $\t_2$ becomes negative in some region.
 This is completely analogous to what happens for D7 brane solutions  in type IIB/F theory, and in that context it is well-known how to
remedy the problem and  construct solutions which do make sense globally \cite{Greene:1989ya}. Such solutions always contain several branes and therefore inevitably break the rotational invariance in the transverse space \cite{Braun:2008ua} and are therefore no longer toric. Let us illustrate this with a simple example, we refer
to \cite{Greene:1989ya},\cite{Bergshoeff:2006jj} for more general solutions.

We describe here a simple global solution   which contains an M2-brane  at $u= 0$. Since we would like $u$ to run over the complex plane and $\t$ to take values
in the fundamental region in the upper half plane, we use  Klein's modular invariant $j$-function to map the fundamental region into the complex plane:
\be
j(\t (u)) = 1 + {1\over u} \label{singleM2}
\ee
Near $u=0 $, this behaves as
\be
\t \sim {1 \over 2 \p i} \ln u + {\rm regular\ terms}
\ee
so that we indeed have monodromy $\t \to \t + 1$ when encircling $u =0$.
Near $u = \infty$ we have
\be
\t \sim i.
\ee
To obtain  a modular invariant metric which is nondegenerate in $u=0$, we choose the function $h$ as:
\be
h =4  \ln \left( \h(\t(u)) u^{-1/24}\right).\label{hsolflatgl}
\ee
The metric then becomes
\be
ds^2_5 =- dt^2 +{  \t_2(u) |\h(\t(u))|^4 \over |u|^{1/6}} du d\bar u+  dw^2 d\bar w^2 \label{globalflatM2}
\ee
It's easy to see that it satisfies the condition (\ref{geq}) for being supersymmetric.
At large $u$, there is a deficit angle of ${2 \p \over 12}$. This solution contains, besides the M2-brane at $u=0$, two elliptic branes
at the points where $\t=i$  and $\t=e^{2 \p i/3}$ with charges $q =1/4$ and $q=1/6$ respectively. Although rotational invariance in the $u$-plane is broken, one can check that the solution indeed takes the toric form
(\ref{flatM2}),(\ref{ellsol}) near the  positions of the various branes.

\subsubsection{Branes in the TN-anti-TN backgound}\label{secprobe}
\begin{figure}
\begin{center}
\begin{picture}(200,120)
\put(0,0){\includegraphics[height=120pt]{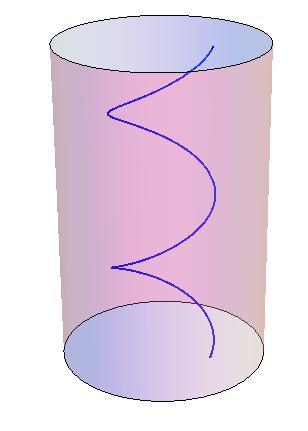}}
\put(30,-10){AdS$_3$}
\put(130,-10){S$^2$}
\put(100,60){$\times$}
\put(110,40){\includegraphics[height=50pt]{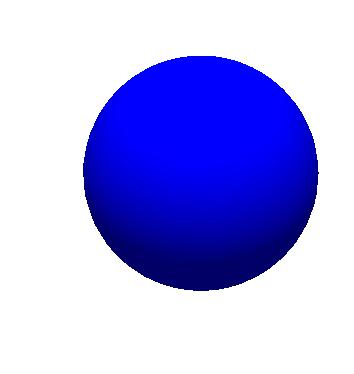}}
\end{picture}\end{center}
\caption{A brane probe at constant $w^1$ in the AdS$_3\times$S$^2$ background wraps the S$^2$ and spirals around in $AdS_3$. }\label{probefig}
\end{figure}
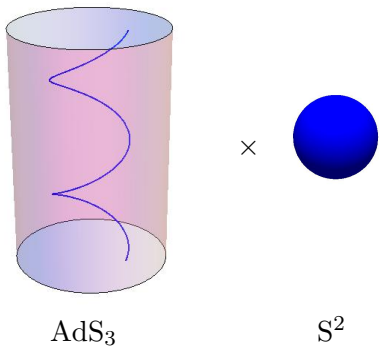

Next we would like to study the  backreaction of an M2-brane (or one of its exotic cousins) in the TN-anti-TN background of example \ref{secads3s2}. As we discussed
there, this background is simply global AdS$_3$ with a round S$^2$ fibered over it. Now consider the submanifold $w^1 =$ constant
in this background, in other words $\r = \r_0, \theta^1 = \theta^1_0$ in the coordinates (\ref{prolspher}). This is a holomorphic surface within the hyperk\"ahler base, and hence a probe M2 or exotic brane placed at this locus will preserve
supersymmetry  \cite{Denef:2007yt}.  Such a BPS brane wraps the S$^2$ and
is static in AdS$_3$  with respect to the time coordinate in (\ref{ads3s2metr}). Nevertheless, because the metric is not static but only stationary in these coordinates, it carries  angular momentum $J_{\theta^1}$ proportional to  $\sinh^2 {\rho_0\over 2}$.  Indeed, in terms of standard
global coordinates such a brane  spirals around at a distance $\r_0$ from the center of  AdS$_3$, see figure \ref{probefig}.
Introducing the brane breaks the  $SL(2,\RR) \times
SL(2,\RR) \times SO(3) $ symmetry of the background down to   the subgroup
\be
U(1) \times U(1) \times SO(3).\label{symmprobe}
\ee

\begin{figure}
\begin{center}
\begin{picture}(200,120)
\put(0,0){\includegraphics[height=120pt]{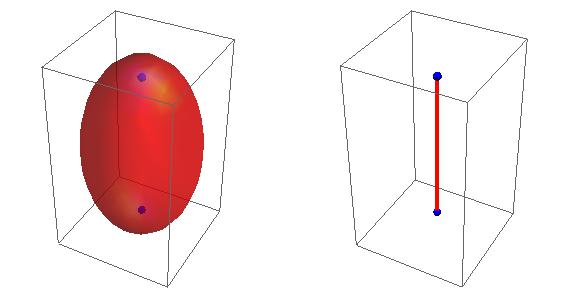}}
\end{picture}\end{center}
\caption{(a) From the 4D point of view, a brane probe at constant $w^1$ is an ellipsoidal D2-brane surroundig a D6 and anti-D6 brane. (b) The brane
probe at the center of $AdS_3$ is a collapsed ellipsoidal D2-brane in 4D.}\label{superegg}
\end{figure}

From the point of view of the 4D theory obtained by dimensionally reducing along $\theta^1$, the M2-brane becomes a D2
brane wrapping an ellipsoidal surface with a D6 and anti-D6 brane at its focal points, see figure \ref{superegg} (a).
 The angular momentum along $\theta^1$
becomes D0-brane charge sourced by worldvolume flux. This brane configuration carries the same charges as a D0-D4 black hole
and was conjectured to play the role of a microstate geometry for this black hole in the black hole deconstruction proposal of \cite{Denef:2007yt}. An
argument from quantizing the probe worldvolume theory also suggests that these microstates are sufficiently typical
to account for the leading contribution to the black hole entropy in a specific large charge limit \cite{Gaiotto:2004ij}.

Let's focus for simplicity on a brane inserted at the `center' of AdS$_3$, meaning at  $\r_0 =0$. Within the 3D base this is the fixed locus
of the Killing vector $\pa_{\theta^1}$, and we know from the discussion in section \ref{intrM2}   that the backreacted solution will still have a toric base with Killing vectors\footnote{Note that the choice $\r =0$ is merely convenience, since because of  homogeneity any of the 
worldlines $\r = \r_0, \theta^1 = \theta^1_0$ discussed above are the fixed locus of some Killing vector  $\pa_{\tilde \theta^1}$, and inserting a brane there will still give rise to a toric configuration with respect to $\pa_{\tilde \theta^1},\pa_{\theta^2}$.} $\pa_{\theta^1},\pa_{\theta^2}$, so let's attempt to find this
solution. From the 4D point of view, the  brane at the center of AdS represents  a collapsed
ellipsoidal brane, see figure \ref{superegg}(b).

Since the background is a solution of the factorized form (\ref{5Dmetrtorambip}-\ref{5Dambisol}) with $\k^2= - 4/P^2, s_1 =0, s_2 =2 ,a^2>0,b=1$, it is natural to also look for the backreacted
solution within the factorized ansatz with these values of the parameters.  Although we don't have a definite proof that the backreacted solution must remain in the factorized form,  we expect that it should since the brane leaves the $SO(3)$ symmetry unbroken and the factorized ansatz naturally leads to solutions with $SO(3)$ invariance as discussed below (\ref{5Dambisol}). Under this assumption, the solution will be of the form (\ref{5Dmetrtorambip}-\ref{5Dambisol}) with the aforementioned values of the parameters and with $\F(x^1)$ a
solution of
\be
\F''+ {4\over P^2} e^{\m - 2 \F} =0\label{liouvsuperegg}
\ee
where $\m(x^1)$ takes one of the three forms given in table \ref{table4} depending on whether the brane is of parabolic (M2), hyperbolic or elliptic type. 
\subsubsection{G\"odel $\times S^2$ and its brane interpretation}\label{Godelsec}
We know already one particular solution to (\ref{liouvsuperegg}), namely (\ref{Gsoltoric}):
\be
e^{2 \F} = {8 e^{3 \m} \over 3 c^2 P^2}.\label{phigodel}
\ee
which we argued  to lead to a highly symmetric 5D solution of the form (\ref{5Dmetrtorambip}-\ref{5Dambisol}). This solution is hence a first  candidate for
a  backreacted brane in AdS$_3\times$S$^2$.   This was the guess made in \cite{Levi:2009az} where, as we shall presently review,  it was also shown to have  several problematic   properties which rule it out  as the backreaction of a single brane. Nevertheless it is clear that the solution must have some interpretation in terms of branes in AdS$3 \times$S$^2$ and we will now give such an interpretation.

Let's first  discuss the solution expected to represent an elliptic brane, where we take
\be
e^\m = \half \sinh 2(1 - q x^1).\label{muell2}
 \ee
 This is perhaps the most symmetric situation since $\pa_{\theta^1}$ generates
 an elliptic isometry within  the $SL(2,\RR)$ symmetry of the 3D metric, so it's natural to first consider the case where $\t$ has an an elliptic monodromy as well.
Plugging into (\ref{phigodel}) leads to a base of the form (\ref{3DmetrsepToda}) which is a toric K\"ahler manifold of the ambipolar type which as far as we know  has not yet appeared in the literature. The  K\"ahler potential  can be read off from (\ref{calkfact}:
\be
\calk = {a P  } \left( \sqrt{1- {2 e^{2 x^2 + 3 \m }\over 3 a^2  q^2 P^2}}- \arctan \sqrt{1- {2 e^{2 x^2 + 3 \m }\over 3 a^2  q^2 P^2}} \right).
\ee
Making a Legendre transformation one finds the symplectic potential
\be
\cals =   y_2\ln {\sqrt{ ( a^2 P^2- y_2^2) (y_1^2 - 9 q^2 y_2^2)^{3/2}}\over q y_2^3}-  a P{\rm arctanh} { a P\over y_2}- {y_1 \over 2 q} {\rm arctanh} { y_1\over 3 q y_2}  - y_2+{y_1\over q}\nonu
\ee
The moment polytope is given by
\be
|y_2| \leq a P, \qquad |y_2|\leq {1\over 3 q} |y_1|
\ee
It is of the same shape as the polytope (\ref{pol66b}) governing the AdS$_3\times$S$^2$ solution depicted in figure \ref{pols}(d); hence this particular polytope admits both  a toric hyperk\"ahler as well as a toric K\"ahler metric.

Switching to  coordinates $\h, \r$
\bea
x^1 &=& {1\over q} \left( 1- \half \ln \tanh {\r \over 2} \right) \\
y_2 &=&  a P \cos \h
\eea
the ambipolar metric on the 4D base is:
\be
ds^2_4 =  -{ a P} \cos \eta  \left( { 3 \over 2 }  \left( d\r^2+ 4 q^2 \sinh^2 \r
  (d \theta^1)^2\right) +d\eta^2 +\tan^2\eta  \left( d\theta^2+3 q \cosh \r
   d \theta^1\right)^2\right)
   \ee
   We note something interesting: taking $\theta^1$ to have period $2\p$ as before, the 3D base has a conical defect singularity for generic charge $q$, but it becomes smooth for
 \be
 q=\half.
\ee
We will take the brane charge to have this value and will comment on its  interpretation below.

    Now we turn to the 5D solution (\ref{5Dmetrtorambip}-\ref{5Dambisol}).
     The parameter $a$   is again fixed by requiring that $d \o$ doesn't have Dirac string singularities, or equivalently, requiring  absence
of CTCs near $x^1 \to \infty$. From (\ref{phigodel})) we read off
the asymptotic behaviour of $\F$, $\F \sim 3 q x^1$, so that in (\ref{distconstrToda}) we have to take $m  = 3 q$ and impose the following relation between the parameters
\be
a  = {p^3 q P^2 \over 8 R}.\label{asolGod}
\ee
The full 5D solution then reads
\bea
\t &=& i \tanh \left(1- \half  w^1 \right)= -{\cos{\theta^1\over 2} + i e^\r  \sin{\theta^1\over 2}\over \sin{\theta^1\over 2} - i e^\r  \cos{\theta^1\over 2}}\\
ds^2_5 &=& {P^2 \over 4} \left({p^3 \over 6}\right)^{2/3} \left[ - \left({3 dt\over R}  + {3\over 2}(\cosh \r -1) d\theta^1\right)^2 + {3 \over 2} \left(d\r^2 +  \sinh^2 \r (d {\theta^1})^2 \right)\right.\nonu
&& \left.+ d\h^2 + \sin^2\h d\left(\theta^2 + {3\over 2} {\theta^1} -  {3 t \over R} \right)^2\right]\\
F^I &=& - { p^I  P\over 2}  \sin \h d\h  \wedge d\left(\theta^2 + {3\over 2} {\theta^1} -  {3 t \over R} \right), \qquad Y^1 = \left({ 6 \over p^3} \right)^{1\over 3} p^I  \label{godsol5d}
\eea
As promised in section \ref{sectoricToda}, this is indeed a `bubbled' solution with a nontrivial S$^2$ supported by flux.

A first unpleasant property of the metric is that it has CTCs. Although the condition  (\ref{asolGod}) removes  CTCs that would otherwise appear near $\r=0$, the full solution develops CTCs for larger values of $\r$ since
\be
g_{\theta^1 \theta^1}\sim  \sinh^2\frac{\rho }{2} (5-\cosh \rho )+\frac{3 \sin ^2\eta }{2}
   \ee
Comparing the first line of (\ref{godsol5d}) with the global AdS$_3$ solution (\ref{ads3s2metr}), we see that the timelike fiber is stretched.
In fact, timelike stretched AdS$_3$  is the most well-known solution with CTCs,  namely the  G\"odel universe \cite{Godel:1949ga}\footnote{To be precise, G\"odel's 4D solution was the product of 3D timelike stretched AdS$_3$ with a line \cite{Rooman:1998xf}.}. Hence our solution represents the 3D G\"odel universe with a spinning
 S$^2$ fibered over it.

Another property which makes the solution unlikely to represent the backreaction of a single brane is that has too much symmetry: the symmetry is
$SL(2,\RR) \times U(1) \times SO(3)$ while from the probe analysis we learned that the  brane should preserve only $U(1) \times U(1) \times SO(3)$. So does this solution actually contain any elliptic branes? At a first glance one might think that the answer is no: for $q=\half$, the accumulated
$SL(2, \ZZ ) $ monodromy under $\theta^1 \to \theta^1 + 2 \p$ is {\bf -1}, i.e. the center  of $SL(2, \ZZ ) $. Since $\t$ itself doesn't feel the
center of $SL(2, \ZZ ) $ and has no monodromy one might be tempted to conclude that for $q=\half$ there is no actual brane source present. This conclusion would be wrong however,  for the same reason that it would be in F-theory, because there are fields in the theory which do feel the center of $SL(2, \ZZ ) $ and have a monodromy. In the
F-theory context, these fields are the NSNS and RR two forms and the object which produces the {\bf -1} monodromy is a bound state of an O7 plane and 4 D7 branes \cite{Sen:1996vd}, while in the present context one can see from (\ref{Fpmmodtransf})
that the  two-forms $\F^+$ and $\F^-$ transform under the center of $SL(2, \ZZ ) $  under which they pick up a sign. Hence we conclude that the G\"odel$\times$S$^2$ solution contains a brane source in the origin. This is not the only brane source present however  since, because the G\"odel universe is spatially homogeneous, we could have actually made the previous argument for any point in the G\"odel universe.
In particular, the solution also contains brane sources on the boundary of the G\"odel universe, which was already observed in \cite{Levi:2009az}.
 
 Hence we conclude that the G\"odel$\times$S$^2$ solution arises from filling  AdS$_3\times$S$^2$ with a congruence of $q= \half$ elliptic branes,
each of them wrapped on the S$^2$ and moving on a rotating worldline as in figure \ref{probefig}.  Indeed, it is well known that the 3D Godel universe
arises from filling  AdS$_3$ with rotating dust, and in the present supersymmetric context the rotating dust consists of  $q= \half$ elliptic branes. Indeed, one can show that the stress tensor of the axidilaton is precisely that of rotating dust \cite{Levi:2009az}.

What if we had started, instead of (\ref{muell2}), from the source $\m$ appropriate for an M2 or hyperbolic brane?  As was discussed in section \ref{sectoricToda}, the solutions in these classes $IP,IH$ are in this case actually related by a holomorphic
 coordinate transformation of $w^1$: they simply correspond to choosing $\pa_{\theta^1}$ to generate a parabolic or hyperbolic
 orbit on the hyperbolic plane instead of an  elliptic one.

We also want to point out an interesting difference between this solution and the brane solutions in flat space of example \ref{secm2flat}. The latter were not globally consistent as $\t_2$ became negative in some region. This is not the case in the current example, since the locus
$\r \to \infty$ where $\t_2$ becomes zero coincides with the boundary of the G\"odel spacetime.

\subsubsection{Outlook: branes in curved backgrounds and black hole deconstruction}\label{secBHdeconstr}
Having established that the G\"odel $\times S^2$ solution does not describe the backreaction of a single brane but rather of a distribution of brane charges,
we now return to the discussion of the backreacted solution of a single  M2-brane in the Taub-NUT-anti-Taub-NUT background which we
initiated  in section \ref{secprobe}.  This solution is expected to be of the form (\ref{5Dmetrtorambip}-\ref{5Dambisol}) and is
 completely determined in terms of a function $\F$ obeying 
the nonlinear ordinary differential equation
 \be
 \F'' + {4 \over P^2} (1- p x^1) e^{- 2 \F }=0\label{M2inTNTNbeq}.
 \ee
 It should obey the subsidiary condition
 \be
 \lim_{p \to 0} e^\F =  2 \sinh {x^1 \over P} \label{subsid}
\ee
which expresses that we should recover the TN-anti-TN background when we turn  off the M2 charge. The special solution (\ref{phigodel})
to (\ref{M2inTNTNbeq}), $e^{2 \F} = {8 (1-p x^1)^3 \over 3 p^2 P^2}$, which gives rise to the G\"odel $\times S^2$ spacetime, doesn't satisfy the subsidiary condition as it becomes singular in the limit $p \to 0$. Hence if the desired  solution
 is to be found in this class, there should exist more general solutions to  (\ref{M2inTNTNbeq}). Naively we indeed expect the general solution to the  second-order ODE (\ref{M2inTNTNbeq}) to depend on two integration constants, which have been fixed to specific values in  (\ref{phigodel}). In support of this we show in appendix \ref{appdefliouv}  that the solution to (\ref{M2inTNTNbeq}) on an interval is fixed uniquely once the values of $\F$ at the two endpoints are given. Another encouraging fact is that, as we saw in section (\ref{septoda}), the generic solution to (\ref{M2inTNTNbeq}) leads to a metric with $U(1) \times U(1) \times SO(3)$
 symmetry, which agrees with the symmetries (\ref{symmprobe}) preserved by the M2-brane probe.
 Unfortunately, the general solution to (\ref{M2inTNTNbeq}) is not known analytically so that in order to make further progress we must
resort to   approximate methods to construct solutions obeying (\ref{M2inTNTNbeq})  and the subsidiary condition (\ref{subsid}). 
We will discuss such methods and the physical properties of the resulting solution in a forthcoming publication \cite{JRDVDB}.

\begin{figure}
\begin{center}
\begin{picture}(200,120)
\put(0,0){\includegraphics[height=120pt]{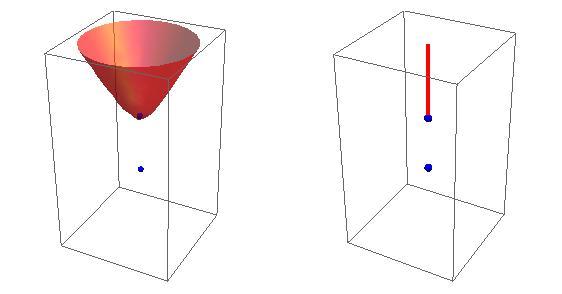}}
\end{picture}\end{center}
\caption{(a) From the 4D point of view, a brane probe at constant $w^1$ in the $\RR \times$Eguchi-Hanson background comes
 from a D2-brane on a single sheet of a hyperboloid in the D6-D6 background. (b) The collapsed  D2 brane at $w^1= - \infty$.}\label{M2EH}
\end{figure}

Let's end by commenting on how similar methods are expected to lead to backreacted M2 brane solutions in the $\RR \times$Eguchi-Hanson and AdS$_2\times$S$^3$ backgrounds. In the $\RR \times$Eguchi-Hanson background discussed in example \ref{secEH}, we can place an M2-brane on the holomorphic surface $w^1=$constant, which  describes the lift to 5D of  a  D2 brane on a single sheet of a hyperboloid in the presence of two D6-branes, see figure \ref{M2EH}(a).
 The limiting case $w^1=-\infty$ is the fixed locus
of $\pa_{\theta^1}$ and inserting the brane there preserves toricity of the base. In the IIA picture, the D2 brane is collapsed and ends on one of the D6 branes, see figure \ref{M2EH}(b). The backreacted solution is  of the type (\ref{EHtype})
 with $s_1 =0, s_2 =2, \k^2 = -4/P^2$ and with $\F (x^1)$ a solution of
 \be
 \F'' - {4 \over P^2}(1- p x^1) e^{- 2 \F }=0\label{M2inTNTNeq}
 \ee
obeying the subsidiary condition
 \be
 \lim_{p \to 0} e^\F =  2 \cosh {x^1 \over P} .\label{subsidEH}
\ee
The same solution to (\ref{M2inTNTNeq},\ref{subsidEH}) also serves to describe a backreacted M2 brane located at $w^1=-\infty$ in the AdS$_2\times$S$^3$ background. In this case the M2 fills AdS$_2$ while tracing out a circle on the  S$^3$. The 5D solution is now of the form (\ref{5Dmetrtorambip}-\ref{5Dambisol}) with $s_1 =0, s_2 =2, \k^2 = 4/P^2, a^2 >0, b=R=0$.

\section*{Acknowledgements}
We dedicate this paper to Iveta and Elis and to Ay\c{c}a and Ozan and thank them for their patience and support.
We would like to  thank Iosif Bena, Hagen Triendl and  Bert Vercnocke for useful discussions.  The research of JR was supported by the Grant Agency of the Czech Republic under the grant
14-31689S.  DVdB is partially supported by TUBITAK grant 113F164 and BAP grant 13B037.

\bigskip

\appendix
\section{Universal hypermultiplet target space}\label{hypermod}
In this Appendix we collect some formulae related to the universal hypermultiplet which will be of use in the main text. We refer to
\cite{Strominger:1997eb} for more details. The target space of the universal hypermultiplet is the  4 real-dimensional quaternionic space ${SU(1,2)\over U(2)}$ on which we choose real coordinates $q^X, X = 1, \ldots 4 $. In terms of the  complex fields $S, C$
\bea
 S &=& q_1 + i q_2  \\
C&=&  q_3 + i q_4
\eea
the metric reads
\be
ds^2 = \psi^1 \bar \psi^1 + \psi^2 \bar \psi^2 \label{hypermetric}
\ee
with
\bea
\psi^1 &=& e^F d C\\
\psi^2 &=& e^{2F} \left( { d S \over 2} - \bar C d \bar C \right)\\
F &=& - \half \log \left( {1\over 2} (S + \bar S)- C \bar C \right)
\eea

The quaternionic structure $J^{i \ Y}_X, i = 1, \ldots 3$ is  explicitly given by
\bea
J^1 &=& {1\over \sqrt{q_1 - q_3^2 - q_4^2}}\left(
\begin{array}{cccc}
 q_4 & q_3 & 0 & \frac{1}{2} \\
 -q_3 & q_4 & -\frac{1}{2} & 0 \\
 -4 q_3 q_4 & 2 \left(q_1-2 q_3^2\right) & -q_4 &
   -q_3 \\
 -2 \left(q_1-2 q_3^2\right) & -4 q_3 q_4 & q_3 &
   -q_4
\end{array}
\right)\nonu
J^2 &=&  {1\over \sqrt{q_1 - q_3^2 - q_4^2}} \left(
\begin{array}{cccc}
 q_3 & -q_4 & \frac{1}{2} & 0 \\
 q_4 & q_3 & 0 & \frac{1}{2} \\
 -2 \left(q_1-2 q_4^2\right) & 4 q_3 q_4 & -q_3 &
   q_4 \\
 -4 q_3 q_4 & -2 \left(q_1-2 q_4^2\right) & -q_4 &
   -q_3
\end{array}
\right)\nonu
J^3 &=& \left(
\begin{array}{cccc}
 0 & -1 & 0 & 0 \\
 1 & 0 & 0 & 0 \\
 4 q_4 & 4 q_3 & 0 & 1 \\
 -4 q_3 & 4 q_4 & -1 & 0
\end{array}
\right)\label{quatJs}
\eea
These satisfy $(J^1)^2=(J^2)^2=(J^2)^2=J^1 J^2 J^3=-1$.

The spin connection splits in $SU(2) \times SU(2)'$ parts: $$\o = - {i \over 2}  A^a \s^a \otimes {\bf 1} +  {\bf 1}\otimes B^a \s^a,$$ with
\bea
A^1 &=& \frac{2 dq_4}{\sqrt{q_1-q_3^2-q_4^2}}\nonu
A^2 &=& \frac{2 dq_3}{\sqrt{q_1-q_3^2-q_4^2}}\nonu
A^3 &=& -\frac{dq_2+2q_4 dq_3-2 q_3
   dq_4}{2 \left(q_1-q_3^2-q_4^2\right)}.\label{su2conn}
\eea

\section{Solving the vector multiplet equations}\label{phase2}
In this Appendix, we derive the general form of solutions to the  equations (\ref{VMeq1}-\ref{VMeq3}) determining how time is fibered over the 4D base as well the fields in  the vector multiplets,
in the case that there is a Killing vector under which $\t$ is invariant. This means we have to solve (\ref{VMeq1}-\ref{VMeq3})
on a 4D base manifold with metric (\ref{4dmetrrot}).

We can treat this system of equations as in the analysis of \cite{Bena:2007ju} which easily generalizes to the
case when the axion-dilaton is turned on. 
To solve the first equation (\ref{VMeq1}), we start from the ansatz
\bea
\Theta^I &=& \left( -2 K^0 \star_3 d\left( { K^I\over K^0} \right) \right)^- \\
&=&  d\left({K^I \over K^0} (d\theta^2 + \chi) \right)-  \star_3 \left( d K^I  + \tilde \t_2 e^{\tilde h_1} s_2 e^{s_2 \Psi} K^0 K^I d y_2 \right).
\eea
Using (\ref{om0eq}) and the property that, for a  one-form $\a$,
\be \star_4 ( \star_3 \a) = \a \wedge {1\over K^0} (d\theta^2 + \chi)\label{starprop}\ee one finds that demanding that $\Theta^I$ is closed is equivalent to
\be
\nabla_3^2 K^I = -s_2  e^{-s_2 \Psi} \pa_{y_2}\left( K^0 K^I e^{s_2 \Psi} \right)\label{eqKuIap}
\ee
where the subscript $\,_3$ means that the covariant derivative is taken with respect to the 3D metric (\ref{3dmetrrot}).
For later use we note that
closed selfdual forms can be constructed starting from the ansatz $\tilde \Theta = \left(  2 K^0 \star_3 dF \right)^+$, leading to the equation
\be
\nabla_3^2 F = s_2  K^0  F_{y_2}\label{closedsd}
\ee

Defining $Z_I = f^{-1} Y_I$, he second equation (\ref{VMeq2}) is equivalent to
\be
\nabla_3^2 Z_I =2 D_{IJK} \nabla_3^i \left( {K^J \over K^0 }\right) \nabla_{3i} \left( {K^K \over K^0 }\right).
\ee
Plugging in the ansatz
\be
Z_I =-  2 K_I +  D_{IJK} {K^J K^K \over K^0}
\ee
and using (\ref{Toda},\ref{eqKuIap}), leads to the following equation for $K_I$:
\be
\nabla_3^2 K_I = -  {s_2 D_{IJK}\over 2}    e^{-s_2 \Psi} \pa_{y_2}\left( K^0 K^I e^{s_2 \Psi} \right)\label{eqKlIap}
\ee

Turning to the last equation (\ref{VMeq3}), we decompose $\xi$ as
\be
\x = \n (d\theta^2 + \chi) + {\o \over 2}
\ee
with $\o$ a one-form on the 3D base.
Contracting the equation with $\pa_{\theta^2}$ gives the equation for $\o$:
\be
\star_3 d \o = - 2 \n d K^0 +2  K^0 d\n - K^0 Z_I d\left( {K^I \over K^0} \right)-2  s_2 \n (K^0)^2 dy_2.
\ee
The remaining equations are equivalent to the integrability condition for this equation and lead to
\be
\nabla_3^2 \n = {1 \over 2 K^0} \nabla_3^i \left( K^0 Z_I \pa_i \left( {K^I \over K^0} \right) \right) + s_2 K^0 \pa_{y_2} \n.
\ee
Plugging in the ansatz
\be
\n = - {K_I K^I \over 2 K^0} + D_{IJK} {K^IK^JK^K\over 6 (K^0)^2} + {K_0\over 2}
\ee
leads to the following equation for $K_0$:
\be
\nabla_3^2 K_0 = s_2 K^0 \pa_{y_2} K_0 - {s_2 } K^I \pa_{y_2} K_I  - {s_2^2 \over 6} D_{IJK} K^IK^JK^K.
\ee
As a check we see from (\ref{closedsd}) that the freedom of adding a closed selfdual part to $d\x$ corresponds to shifting $K_0$ by a solution of the homogeneous equation. Putting this all together leads to the general form of the solution (\ref{fluxsol1}-\ref{fssol}) and the differential equations (\ref{eqKu0}-\ref{omeq}).

\section{Boundary value problem for the deformed Liouville equation}\label{appdefliouv}
Here we will show, following \cite{trojanov}, that the solution of the deformed Liouville equation (\ref{Liouv}) with $\k^2 <0$ on an interval is uniquely determined
by by specifying its values at the endpoints. Setting $\k^2=-1, s_2=0$ by redefining $\m$, the deformed Liouville equation (\ref{Liouvtoric}) reads
\be
\F'' +  e^{\m -2 \F} =0
\ee
which we consider on some interval $\cald$ of the real line.
Suppose we have two solutions $\F_1,\F_2$, then it follows that their difference $w = \F_1 - \F_2$ satisfies
\be
w'' =  e^{\m -2 \F_1} w (e^{2w}-1)\geq 0
\ee
Multiplying with $w$ and integrating we get
\be
- \int_\cald dx (w')^2 + ( w w')_{|\d \cald} = \int_\cald  e^{\m-2 \F_1}w (e^{2w}-1) \geq 0
\ee
If the boundary term vanishes, i.e. if we fix either $\F$ or $\F'$ on the boundary, we can conclude that $w=0$ and the solution is unique.
When the boundary behaviour is not fixed the solution depends on two integration constants as one would naively expect.

\end{document}